\documentclass{emulateapj}
\usepackage{apjfonts}

\begin{document}

\lefthead{Chandra Observations of IGR Sources}
\righthead{Tomsick et al.}

\submitted{Accepted by the Astrophysical Journal}

\def\lsim{\mathrel{\lower .85ex\hbox{\rlap{$\sim$}\raise
.95ex\hbox{$<$} }}}
\def\gsim{\mathrel{\lower .80ex\hbox{\rlap{$\sim$}\raise
.90ex\hbox{$>$} }}}

\title{{\em Chandra} Localizations and Spectra of {\em INTEGRAL}
Sources in the Galactic Plane}

\author{John A. Tomsick\altaffilmark{1},
Sylvain Chaty\altaffilmark{2},
Jerome Rodriguez\altaffilmark{2},
Roland Walter\altaffilmark{3},
Philip Kaaret\altaffilmark{4}}

\altaffiltext{1}{Space Sciences Laboratory, 7 Gauss Way, 
University of California, Berkeley, CA 94720-7450, USA
(e-mail: jtomsick@ssl.berkeley.edu)}

\altaffiltext{2}{AIM - Astrophysique Interactions Multi-\'echelles
(UMR 7158 CEA/CNRS/Universit\'e Paris 7 Denis Diderot),
CEA Saclay, DSM/IRFU/Service d'Astrophysique, B\^at. 709,
L'Orme des Merisiers, FR-91 191 Gif-sur-Yvette Cedex, France}

\altaffiltext{3}{INTEGRAL Science Data Centre, Observatoire
de Gen\'eve, Universit\'e de Gen\'eve, Chemin d'Ecogia, 16, 
1290 Versoix, Switzerland}

\altaffiltext{4}{Department of Physics and Astronomy, University of
Iowa, Iowa City, IA 52242, USA}

\begin{abstract}

We report on the results of observations of hard X-ray sources in the 
Galactic plane with the {\em Chandra X-ray Observatory}.  The hard X-ray 
``IGR'' sources were discovered by the {\em INTEGRAL} satellite, and 
the goals of the {\em Chandra} observations are to provide sub-arcsecond
localizations to obtain optical and infrared counterparts and to provide 
constraints on their 0.3--10 keV spectra.  We obtained relatively short, 
$\sim$5 ks, observations for 20 IGR sources and find a bright {\em Chandra} 
source in {\em INTEGRAL} error circles in 12 cases.  In 11 of these cases, 
a cross-correlation with optical and/or infrared source catalogs yields a 
counterpart, and the range of $J$-band magnitudes is 8.1--16.4.  Also, in 
4 cases, the {\em Chandra} X-ray spectra show evidence for absorbing 
material surrounding the compact object with a column density of local 
material in excess of $5\times 10^{22}$~cm$^{-2}$.  We confirm that 
IGR~J00234+6141 is a Cataclysmic Variable and IGR~J14515--5542 is an Active 
Galactic Nucleus (AGN).  We also confirm that IGR~J06074+2205, IGR~J10101--5645, 
IGR~J11305--6256, and IGR~J17200--3116 are High Mass X-ray Binaries (HMXBs).  
Our results (along with follow-up optical spectroscopy reported elsewhere) 
indicate that IGR~J11435--6109 is an HMXB and IGR~J18259--0706 is an AGN.
We find that IGR~J09026--4812, IGR~J18214--1318, and IGR~J18325--0756 may 
be HMXBs.  In cases where we do not find a {\em Chandra} counterpart, the
flux upper limits place interesting constraints on the luminosities of 
black hole and neutron star X-ray transients in quiescence.

\end{abstract}


\section{Introduction}

The hard X-ray imaging of the Galactic plane by the {\em INTErnational 
Gamma-Ray Astrophysics Laboratory (INTEGRAL)} satellite is uncovering a 
large number of new or previously poorly studied ``IGR" sources.  In 
the most recent comprehensive paper on {\em INTEGRAL} sources, 500
sources had been detected by {\em INTEGRAL} in the 20--40 keV band, 
including 214 IGR sources \citep{bodaghee07}.  While {\em INTEGRAL} 
provides hard X-ray images of unprecedented quality, it still only 
localizes sources to $1^{\prime}$--$5^{\prime}$, which is not nearly 
adequate for reliably finding optical or IR counterparts.  However, 
a relatively large fraction of the IGR sources are persistent, making
it possible to detect the sources in the $\sim$1--10 keV band with 
{\em Chandra}, {\em XMM-Newton}, or {\em Swift}, vastly improving their 
localizations even with short observations with these satellites 
\citep{tomsick06,rodriguez03,rtc08}.

The group of IGR sources that have been identified include 49 Active 
Galactic Nuclei (AGN), 32 High-Mass X-ray Binaries (HMXBs), 9 Low-Mass 
X-ray Binaries, 9 Cataclysmic Variables (CVs), and smaller numbers of 
other source types \citep{bird06,bodaghee07}.  Probably the biggest 
surprise is the large number of HMXBs as well as the properties of these 
systems.  Many of the IGR HMXBs have large levels of intrinsic (i.e., 
local) absorption with $N_{\rm H}\sim 10^{23-24}$ cm$^{-2}$ 
\citep[e.g.,][]{walter03,mg03,rodriguez03,combi04,walter04c,beckmann05,walter06}, 
and there is evidence in several cases that a strong stellar equatorial outflow
is responsible for the absorption \citep{fc04,chaty08}.  In addition 
to the high column densities, some members of the group of IGR HMXBs 
exhibit other extreme properties, including the high amplitude X-ray 
flaring of the Supergiant Fast X-ray Transients
\citep[SFXTs,][]{negueruela06,sguera06,wz07} and neutron stars 
with very long, $\sim$1000--6000~s, spin periods 
\citep{patel04,lutovinov05b}, which may, in at least some cases, 
be due to very high neutron star magnetic fields \citep{patel07}.

In order to determine the nature of more of the unclassified IGR
sources, we have been obtaining observations of IGR sources in the 
Galactic plane with the {\em Chandra X-ray Observatory}.  We observed 
4 sources during {\em Chandra} observing cycle 6, obtained sub-arcsecond 
localizations for all 4, and found that 2 of these sources are CVs, 
1 is an HMXB, and 1 is a likely HMXB \citep{tomsick06,masetti06v,chaty08}.  
When we selected targets for our cycle 8 proposal (in 2006 March), we 
considered all the IGR sources that had been detected at that time (now, 
the number has nearly doubled, and we are observing many of the recently 
detected sources in cycle 9).  As we are interested in Galactic sources, 
we selected sources that are within $5^{\circ}$ of the Galactic plane.
We also eliminated sources from our list that were considered to be
likely Active Galactic Nuclei (AGN), and this left us with a list of 
20 targets.  

Between 2006 December and 2008 February, we obtained cycle 8 {\em Chandra} 
observations of these 20 IGR sources.  The goals of the observations are 
to provide a sub-arcsecond position for each source to facilitate an 
optical/IR identification and to obtain a soft X-ray spectrum.  In the 
following, we describe the {\em Chandra} observations and our analysis 
of these data in \S$2$.  For the cases where we obtained sub-arcsecond 
{\em Chandra} positions, we report on the results of our search for optical 
and IR counterparts in \S$3$.  Then, in \S$4$, we describe the implications 
for each source and give our conclusions in \S$5$.

\section{{\em Chandra} Observations and Analysis}

\subsection{{\em Chandra} Observations}

Between 2006 December and 2008 February, we obtained cycle 8 {\em Chandra}
observations of 20 IGR sources.  The ObsIDs for these $\sim$5~ks ``snapshot''
observations taken with the Advanced CCD Imaging Spectrometer 
\citep[ACIS,][]{garmire03} are 7518--7535, 7561, and 7562, and the IGR
source names and more details on the observations can be found in 
Table~\ref{tab:obs}.  For each observation, we started with a ``level 1'' 
event list produced via the {\em Chandra} data pipeline.  We downloaded 
the most recent level 1 data products as of 2008 March after these were 
processed at the {\em Chandra} X-ray Center with pipeline (``ASCDS'') 
versions between 7.6.9 and 7.6.11.1.  We performed all further data 
processing locally using the {\em Chandra} Interactive Analysis of 
Observations (CIAO) version 4.0 software and Calibration Data Base (CALDB) 
version 3.4.2.  For each observation, we used the CIAO routine 
{\tt acis\_process\_events} to obtain a ``level 2'' event list and 
image.  We note that while a level 2 file is produced in the pipeline
data, the primary reason that we re-produced the file is so that we
would be using the most recent time-dependent gain information.

Nineteen of the observations were carried out with ACIS-S, while the
observation of IGR~J11305--6256 (ObsID 7527) was carried out with the
ACIS-I to obtain a 16-by-16 arcminute$^{2}$ continuous field-of-view 
(FOV) rather than the 8-by-32 arcminute$^{2}$ continuous FOV for
ACIS-S with 4 CCD detectors activated.  Although the FOV is more 
favorable for ACIS-I, ACIS-S was our first choice because of its 
better soft X-ray response.  The radii of the {\em INTEGRAL} 
90\% confidence error circles range from $1^{\prime}.1$ to 
$4^{\prime}.8$ \citep{bird07,bird06,chenevez04}, and our {\em Chandra}
observations covered the entire {\em INTEGRAL} error circle in all
cases except one, IGR~J00234+6141 (ObsID 7424).  Although the 
{\em Chandra} FOV only covered about 80\% of the IGR~J00234+6141
error circle, this was a case where we detected a bright {\em Chandra} 
source in the {\em INTEGRAL} error circle.

\subsection{Search for Sources and X-ray Identifications}

While several of the observations show bright {\em Chandra} sources
in their corresponding {\em INTEGRAL} error circles that are obvious
likely counterparts, we performed a formal search for sources using 
CIAO program {\tt wavdetect} \citep{freeman02}.  In all cases
(except IGR~J00234+6141 as described above), the search covered the 
full {\em INTEGRAL} error circle, and the search typically extended
over the ACIS chip containing the observation aimpoint (ACIS-S3).
In a few cases, the {\em INTEGRAL} error circle extended to the
adjacent chip (ACIS-S2 or ACIS-S4), and we extended the search to 
part or all of the adjacent chip.  For the search, we used the
0.3--10~keV images and a detection threshold of $10^{-6}$, which
corresponds to the likely detection of 1 spurious source per
ACIS chip (with 1024-by-1024 pixels per chip).

The number of {\em Chandra} sources detected in each {\em INTEGRAL} 
error circle varies widely from zero to 12 sources.  In total, 65
sources are detected, and the sum of the error circle areas is
475 arcmin$^{2}$ for an average source density of 
0.137 sources/arcmin$^{2}$.  Thus, one expects 0.52 sources for
the smallest $1^{\prime}.1$ error circle (we detect zero) and 
9.4 sources for the largest $4^{\prime}.8$ error circle (we
detect 12).  Although one does not necessarily expect the source
density to be uniform for different pointing directions through
the Galaxy, our numbers of detected sources do not require large
variations in the source density.

\begin{figure}
\plotone{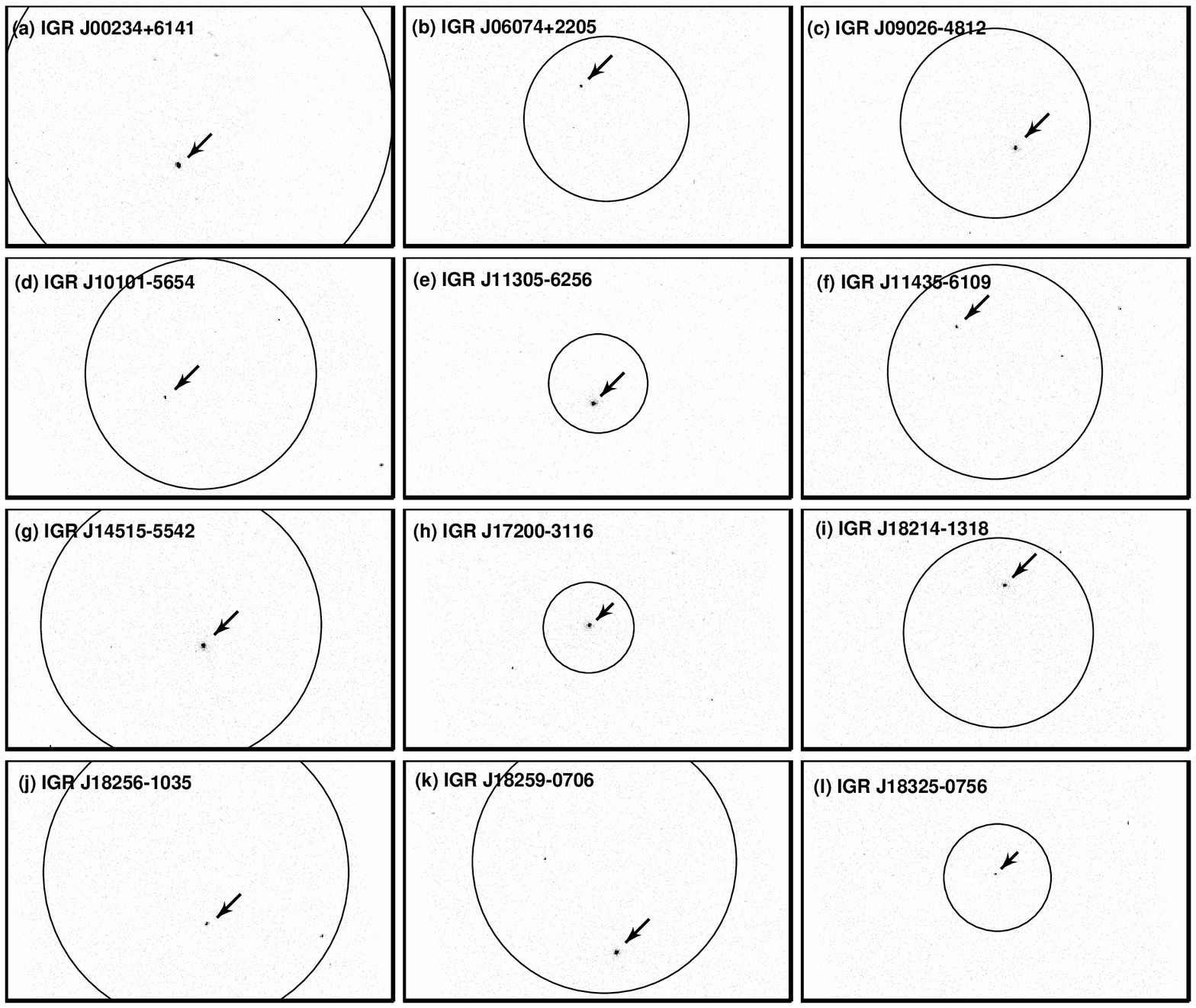}
\vspace{1cm}
\caption{{\em Chandra} 0.3--10 keV images of the 12 IGR-source fields
taken with the ACIS instrument.  The 90\% confidence {\em INTEGRAL}
error circles are shown in each case, and their radii are 
(a) $4^{\prime}.8$, (b) $2^{\prime}$, (c) $2^{\prime}$, (d) $2^{\prime}.8$, 
(e) $1^{\prime}.2$, (f) $2^{\prime}.6$, (g) $3^{\prime}.4$, (h) $1^{\prime}$, 
(i) $2^{\prime}.3$, (j) $3^{\prime}.7$, (k) $3^{\prime}.2$, and (l) $1^{\prime}.3$.
The images have been re-binned by a factor of 4, giving $2^{\prime\prime}$ pixels.
The images are oriented so that North is up and East is to the left.  Arrows point 
to the brightest {\em Chandra} source detected in each observation, and these are 
very likely the soft X-ray counterparts to the IGR sources.  For the brightest
and faintest sources, respectively, 2,200 and 30 counts were detected by 
{\em Chandra}.\label{fig:images}}
\end{figure}

The count rates for these 65 sources also vary widely from 
$<$$1\times 10^{-3}$~c/s ($<$5 counts) to nearly 0.5~c/s ($\sim$2,500 
counts).  The median number of counts per source is 9.6, while the mean
is 141.8, indicating that there is a population of anomalously bright 
sources in at least some of these pointings.  In fact, in 11 of our 
observations, there are single bright sources with count rates between
0.055 and 0.446~c/s.  In our previous {\em Chandra} study of four IGR 
sources \citep{tomsick06}, we measured ACIS count rates between 0.039 
and 0.18~c/s, so the rates for the 11 bright sources mentioned above
are in-line with what we expect if these are the {\em Chandra} 
counterparts to the IGR sources.  The {\em Chandra} positions and ACIS 
count rates for these sources are listed in Table~\ref{tab:sources}.  
There is one other case where we also believe there is strong evidence 
that our {\em Chandra} observations have uncovered the correct counterpart.  
For IGR~J18325--0756, we detect only one source in the $1^{\prime}.3$
{\em INTEGRAL} error circle.  Although this is only a 30 count source 
(0.0059~c/s), it is very close to the center of a relatively small
{\em INTEGRAL} error circle, and we have listed it in Table~\ref{tab:sources}
as a likely counterpart to IGR~J18325--0756.  The {\em Chandra} images for 
these 12 sources are shown in Figure~\ref{fig:images}.

The other 8 observations do not yield clear counterparts.  In 3 cases 
(IGR~J18539+0727, IGR~J18193--2542, and IGR~J17331--2406), no 
{\em Chandra} sources are detected in the {\em INTEGRAL} error circles.  
In the other 5 cases (IGR~J01363+6610, IGR~J17407--2808, IGR~J17445--2747, 
IGR~J17285--2922, and IGR~J17331--2406), multiple (between 2 and 7) 
{\em Chandra} sources are detected, but, in each case, no one source is 
anomalously bright.  For the {\em Chandra} sources detected in these 
observations, the numbers of ACIS counts collected are between 4 and 16.
Although these faint {\em Chandra} sources may possibly be counterparts
to the IGR sources, we do not have any basis for picking which {\em Chandra}
source is the correct counterpart.  A list of {\em Chandra} sources
detected in the {\em INTEGRAL} error circles for these five observations
is provided in Table~\ref{tab:list}.  There are at least 3 possibilities 
why we are not able to find unique counterparts in these 8 cases:  The IGR
sources may be variable or transient; the IGR sources may be highly absorbed;
or the IGR sources may lie outside of the 90\% confidence {\em INTEGRAL} 
error circles we have been considering.  In \S$4.8$, we discuss each of the 
8 cases where we are not able to determine a unique counterpart in more detail.

For the observations without clear counterparts, we also inspected the 
portion of the ACIS FOV outside of the {\em INTEGRAL} error circles, 
and there are notable sources in 2 cases.  For IGR~J17407--2808, the 
{\em INTEGRAL} error circle radius is $4^{\prime}.2$, and we detect a very 
bright, $\sim$5,400 count source $6^{\prime}.0$ from the center of the 
{\em INTEGRAL} error circle.  The position of this source is
R.A. = $17^{\rm h}40^{\rm m}42^{\rm s}\!.84$, 
Decl. = --$28^{\circ}18^{\prime}08^{\prime\prime}\!.5$
(equinox 2000.0, 90\% confidence uncertainty = $0^{\prime\prime}.64$), 
which is consistent with the location of the neutron star LMXB
SLX~1737--282 \citep{tomsick07_atel2}.  It is clear that this 
{\em Chandra} source, CXOU~J174042.8--281808 is SLX~1737--282, 
but it is doubtful that SLX~1737--282 and IGR~J17407--2808 are
the same source.  The other case where there is a possible 
counterpart outside of the {\em INTEGRAL} error circle is
IGR~J17445--2747. While the {\em INTEGRAL} position is good
to $2^{\prime}.3$, we find a 211 count source that is 
$4^{\prime}.4$ away from the best {\em INTEGRAL} position at 
R.A. = $17^{\rm h}44^{\rm m}46^{\rm s}\!.02$, 
Decl. = --$27^{\circ}47^{\prime}32^{\prime\prime}\!.7$
(equinox 2000.0, 90\% confidence undertainty = $0^{\prime\prime}.64$).
This source, CXOU 174446.0--274732, is on the edge of the 
$5^{\prime\prime}.1$ {\em Swift} error circle for a source that
was a suggested as a likely counterpart for IGR~J17445--2747
\citep{landi07_atel}.  However, since both sources are well outside
the {\em INTEGRAL} error circle, it is not clear that either of
them is associated with IGR~J17445--2747.

\subsection{Spectral Analysis}

We extracted 0.3--10~keV ACIS spectra for each of the 12 {\em Chandra}
sources that have likely associations with IGR sources using the CIAO 
tool {\tt psextract}.  We extracted source photons from a circular region
centered on each source with radii of between $5^{\prime\prime}$ and
$7^{\prime\prime}.5$.  The somewhat larger radii are necessary for 
the cases where the {\em Chandra} sources are farther off-axis causing
the source photons to be spread over more pixels.  We also extracted
background spectra from an annulus with an inner radius of 
$10^{\prime\prime}$ and an outer radius of $50^{\prime\prime}$ centered 
on the source.  We used the CIAO tools {\tt mkacisrmf} and {\tt mkarf}
to produce the ACIS response matrices.  While we produced spectra
re-binned to 11 spectral bins in order to provide an opportunity to
look for spectral features, we make use of both binned and unbinned
spectra in this work.

We fitted the binned 0.3--10~keV ACIS spectra using the XSPEC v12 software 
with an absorbed power-law model using $\chi^{2}$ statistics. To account 
for absorption, we used the photoelectric absorption cross sections from 
\cite{bm92} and elemental abundances from \cite{wam00}, which correspond 
to the estimated abundances for the interstellar medium.  In 5 cases, we 
obtained very poor fits with reduced-$\chi^{2}$ values of between 4 and 10 
for 9 degrees of freedom (dof), and given the relatively high count rates 
in these cases (between 0.197 and 0.235 c/s) as well as the fact that the 
strongest residuals are at the high energy end of these spectra (e.g., 
5--10 keV), we suspect that photon pile-up is the primary cause of the poor 
fits.  We checked for pile-up for all 12 sources by looking at the number 
of counts detected above 10 keV, where the ACIS effective area is too 
small to expect any source photons to be collected.  In the 5 cases where 
we suspect pile-up due to poor fits (IGR~J09026--4812, IGR~J11305--6256, 
IGR~J14515--5542, IGR~J17200--3116, and IGR~J18214--1318), between 9 and 
43 counts are detected above 10 keV compared to between 0 and 2 counts in 
the other 7 cases.  This confirms that we must correct for pile-up for the 
spectra of the 5 sources listed above.  Finally, we note that the count 
rates for IGR~J00234+6141 and IGR~J18259--0706 are even higher, at levels
of 0.446 and 0.487 c/s, respectively.  However, these two sources are also
the farthest off-axis at 4--5$^{\prime}$.  With the larger off-axis point 
spread function the source counts are spread over more pixels, and we do 
not see any evidence for significant pile-up in these cases.

\begin{figure}
\plotone{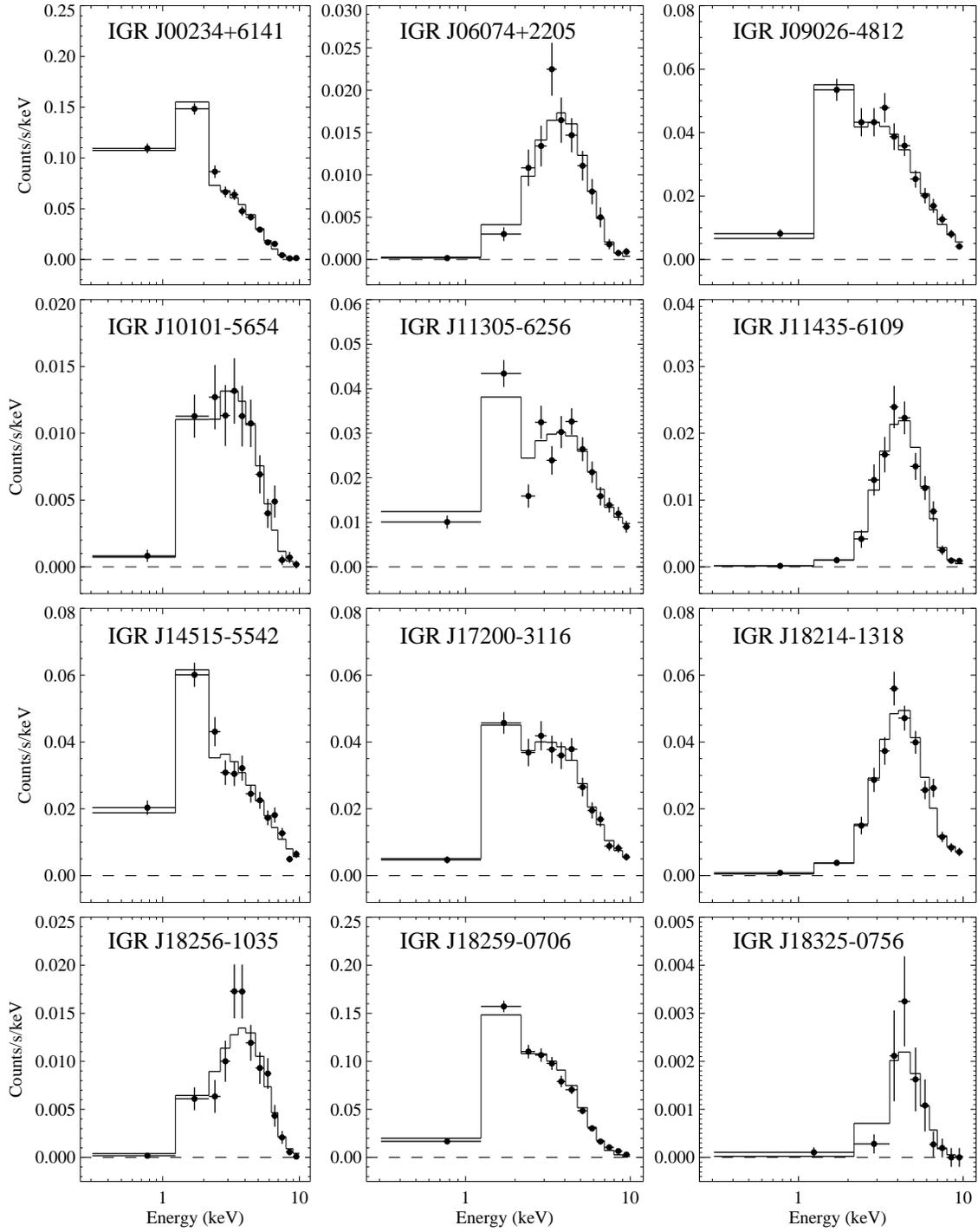}
\vspace{1cm}
\caption{{\em Chandra} 0.3--10 keV spectra of the 12 IGR sources
taken with the ACIS instrument.  The spectra are fitted with an absorbed 
power-law model, and photon pile-up must be included in the modeling for
5 of the sources.  The model (and pile-up) parameters are given in
Table~\ref{tab:spectra}.\label{fig:spectra}}
\end{figure}

We re-fitted the 12 spectra using the same absorbed power-law model, 
but with the addition of the \cite{davis01} pile-up model in the 5 cases
where pile-up is significant.  Also, for these fits, we used the unbinned
spectra and found the best fit parameters using the Cash statistic.  The
results are given in Table~\ref{tab:spectra}, and the parameters indicate
that these sources do indeed have hard spectra, which provides further
confirmation that these are the counterparts to the IGR sources.  For 
11 sources (all except IGR J18325--0756), the best fit values for the 
power-law photon index are between $\Gamma = 0.1$ and 1.5, which is 
inconsistent with a blackbody spectrum and indicates a hard component 
to the spectrum.  For IGR J18325--0756, only 30 counts were collected, 
and the spectral shape is not well-constrained as we obtain 
$\Gamma = 2.8^{+5.4}_{-2.7}$ (90\% confidence).  However, we also obtain 
a column density of $N_{\rm H} = (3.4^{+4.4}_{-2.3})\times 10^{23}$ cm$^{-2}$, 
suggesting that this source may have local absorption like many of the IGR 
sources.  In fact, several of the sources show evidence for local absorption 
with values of $N_{\rm H}$ that are significantly higher than the inferred
column densities ($N_{\rm H}$ and $N_{\rm H_{2}}$) through the Galaxy along 
their lines of sight (see Table~\ref{tab:spectra}).  In addition to 
IGR~J18325--0756, sources IGR~J06074+2205, IGR~J11435--6109, and IGR~J18214--1318 
have $N_{\rm H}$ values of $(7.2^{+2.5}_{-2.0})\times 10^{22}$, 
$(1.5^{+0.5}_{-0.4})\times 10^{23}$, and $(1.17^{+0.30}_{-0.27})\times 10^{23}$
cm$^{-2}$, while the Galactic $N_{\rm H}$ values for these 3 sources are 
between $6.1\times 10^{21}$ and $1.6\times 10^{22}$ cm$^{-2}$ \citep{dl90}
and the Galactic $N_{\rm H_{2}}$ values are between $1\times 10^{20}$ and
$4\times 10^{21}$~cm$^{-2}$.  The {\em Chandra} spectra are shown in 
Figure~\ref{fig:spectra}, and one can see the higher levels of absorption for 
these 4 sources.

To estimate possible levels of long-term variability for these 12 sources,
we used the {\em Chandra} spectral parameters from Table~\ref{tab:spectra}
to compare the flux levels we are seeing with {\em Chandra} to the flux 
levels measured in the 20--40~keV band by {\em INTEGRAL} as reported in
\cite{bird07}.  As the energy bands do not overlap, we compared the fluxes 
by extrapolating the {\em Chandra} power-law spectra into the 20--40~keV band.
We note that there is one exception, IGR~J06074+2205, which was only detected
by {\em INTEGRAL} in the 3--10 keV band, so we performed the comparison in that
energy band.  Table~\ref{tab:comparison} shows the results of the comparisons, 
including a calculation of the ratio of the extrapolated {\em Chandra} flux to 
the {\em INTEGRAL} flux.  The three sources with the lowest values, IGR~J06074+2205, 
IGR~J10101--5645, and IGR~J18325--0756, have ratios of $0.021\pm 0.009$, 
$0.05\pm 0.04$, and $<$0.03, respectively, indicating high levels of variability, 
and these sources might be described as transient.  For the other 9 sources, the 
best values for the ratios are between 0.17 and 1.7, although it should be noted 
that most have relatively large error bars and also that determining whether a
source is variable depends on our assumption about the power-law extending into
the 20--40~keV band.

\section{Optical/IR Identifications}

\begin{figure}
\plotone{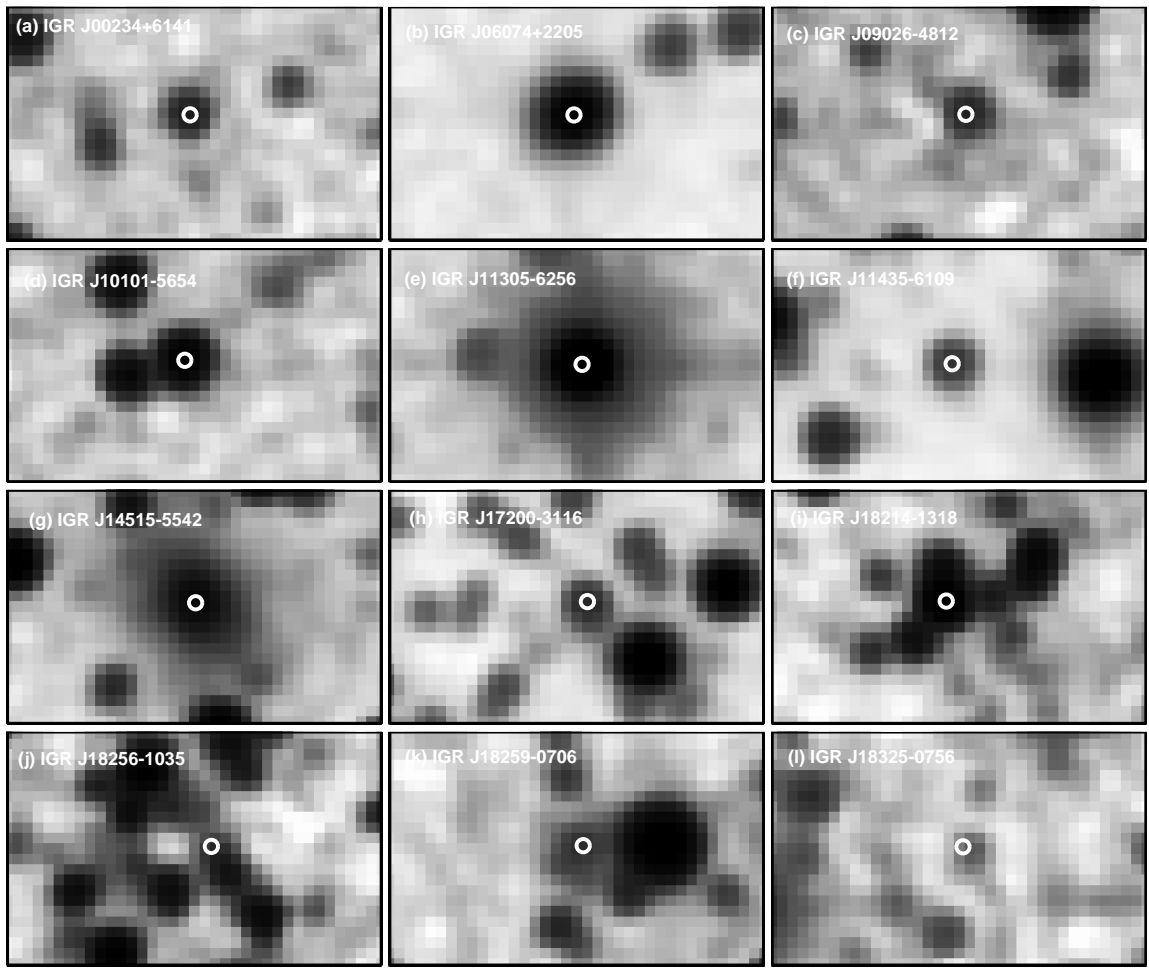}
\vspace{1cm}
\caption{2MASS $J$-band images of the 12 IGR-source fields.  The white
circle marks the 90\% confidence $0^{\prime\prime}.64$ {\em Chandra}
error circle.  The images are oriented so that North is up and East
is to the left.  The pixel sizes for the 2MASS images are
$1^{\prime\prime}$, and the size of each image is $34^{\prime\prime}$
in the East-West direction and $22^{\prime\prime}$ in the North-South
direction.\label{fig:images_2mass}}
\end{figure}

With the {\em Chandra} positions for the 12 IGR sources, we searched
for optical/IR counterparts in the following catalogs:  the 2 Micron
All-Sky Survey (2MASS); the Deep Near Infrared Survey of the Southern 
Sky (DENIS); and the United States Naval Observatory (USNO-B1.0), and 
the results of the search are given in Table~\ref{tab:oir}.  For
each source, the table lists the 2MASS source that is closest to 
the {\em Chandra} position.  Given the {\em Chandra} position 
uncertainties of $0^{\prime\prime}.64$ and the 2MASS position 
uncertainties of $0^{\prime\prime}.2$, only sources with 
{\em Chandra}/2MASS separations less than $\sim$$0^{\prime\prime}.7$
should be considered as possible or likely counterparts.  In 11
cases, there are 2MASS sources with positions consistent with the
{\em Chandra} positions, and we consider these associations to be
likely.  For IGR~J18256--1035, the {\em Chandra}/2MASS separation is 
$5^{\prime\prime}.7$, indicating that these two sources are not
associated.  Figure~\ref{fig:images_2mass} shows 2MASS $J$-band
images with the {\em Chandra} positions marked for all 12 fields.
In the case of IGR~J18256--1035 (Figure~\ref{fig:images_2mass}j), 
it is apparent that the region is very crowded, and the {\em Chandra} 
source falls in an area where several IR sources are blended together.  

The 11 sources with likely 2MASS counterparts have a wide range
of brightnesses from $J = 8.0$ to $J = 16.4$.  The brightest of
the sources is IGR~J11305--6256, and this source appears in the
2MASS, DENIS, and USNO-B1.0 catalogs at optical and IR magnitudes
near 8.  As discussed below, the very low extinction ($J$--$K_{s} 
= 0.04\pm 0.04$) likely indicates that the source is relatively 
nearby.  On the other hand, IGR~J09026--4812, which has
$J = 15.57\pm 0.08$, shows a highly reddened spectral energy 
distribution with $J$--$K_{s} = 2.88\pm 0.09$.  Either extinction
or a very red intrinsic spectrum (more likely the former) explain
why this source is not detected in the optical (down to the 
USNO-B1.0 limit).  Although many of the regions are too crowded
to comment on whether the sources are extended in the 2MASS 
images or not, Figure~\ref{fig:images_2mass}g shows that 
IGR~J14515--5542 is clearly extended.  Our {\em Chandra} position
confirms that this source is an AGN as suggested by 
\cite{masetti06v}.  An inspection of the other sources shown in 
Figure~\ref{fig:images_2mass} does not suggest that they are 
extended, and many of the sources that are relatively isolated
appear to be point-like.

While we believe that the association of IGR~J18325--0756 with
2MASS~J18322828--0756420 is likely, this is the faintest IR source
in our sample ($J = 16.4$) as well as having the largest 
{\em Chandra}/2MASS separation ($0^{\prime\prime}.63$).  Thus, in
this work, we include IR observations of the IGR~J18325--0756 field 
taken at the European Southern Observatory's 3.5 meter New Technology 
Telescope (ESO's NTT) at La Silla Observatory.  We performed NIR 
photometry in $J$, $H$ and $K_{s}$ bands of IGR~J18325--0756 on 
2004 July 11 with the spectro-imager SofI (Son of Isaac).  We used 
the large field of SofI's detector, giving an image scale of 
$0^{\prime\prime}.288$/pixel and a field of view of 4.92 arcmin$^{2}$.  
We obtained the photometric observations for each filter following the 
standard jitter procedure.  This consists of a repeating pattern of 
9 different $30^{\prime\prime}$ offset positions, allowing us to 
cleanly subtract the sky emission in NIR.  The integration time was 
50~s for each individual exposure, giving a total exposure time of 
450~s.  We observed three photometric standard stars from the faint 
NIR standard star catalog of \cite{persson98}: [PMK98] 9157, 
[PMK98] 9172, [PMK98] 9181.

\begin{figure}
\plotone{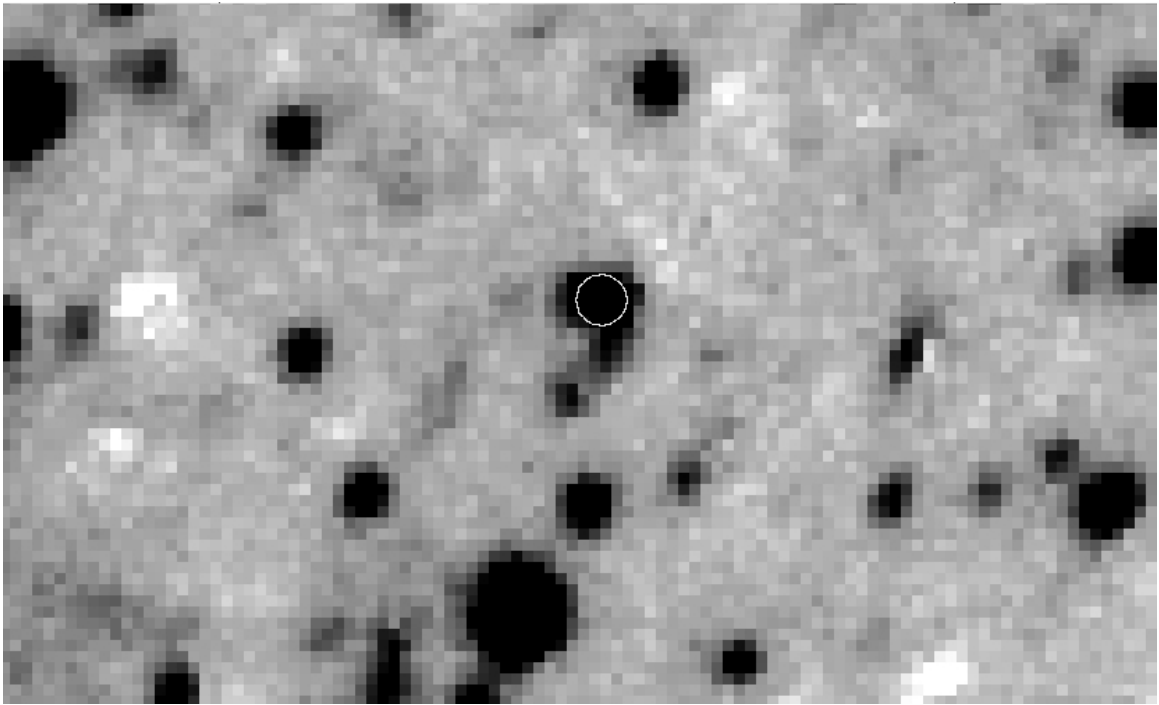}
\vspace{1cm}
\caption{$K_{s}$-band image of the IGR~J18325--0756 field taken with
the SofI instrument on the 3.5 meter New Technology Telescope (NTT).  
The image is oriented so that North is up and East is to the left.  
The pixel size is $0^{\prime\prime}.288$.  The white circle marks
the position of the {\em Chandra} source CXOU~J183228.3--075641, 
and its position is consistent with that of 2MASS~J18322828--0756420, 
which has a $K_{s}$-magnitude of $14.27\pm 0.08$.\label{fig:image_j18325}}
\end{figure}

We used the Image Reduction and Analysis Facility (IRAF) suite to 
perform data reduction, carrying out standard procedures of NIR image 
reduction, including flat-fielding and NIR sky subtraction.  We 
performed accurate astrometry using all stars from the 2MASS catalog 
present in the SofI field, amounting to $\sim$1000 2MASS objects, and
the rms of astrometry fit is $<$$0^{\prime\prime}.5$.  The $K_{s}$-band
image is shown Figure~\ref{fig:image_j18325}, and the {\em Chandra}
position is consistent with the position of the IR source.  We carried 
out aperture photometry, and we then transformed instrumental magnitudes 
into apparent magnitudes using the standard relation: 
$mag_{app} = mag_{inst} - Zp - ext \times AM$, where $mag_{app}$ and 
$mag_{inst}$ are the apparent and instrumental magnitudes, $Zp$ is the 
zero-point, $ext$ is the extinction, and $AM$ is the airmass.  For the 
extinction, we used $ext_\mathrm{J} = 0.06$, $ext_\mathrm{H} = 0.04$, 
and $ext_\mathrm{Ks} = 0.10$, typical of La Silla Observatory.  The 
apparent magnitudes are $J = 16.26\pm 0.04$, $H = 15.13\pm 0.06$, and
$K_{s} = 14.27\pm 0.08$, which are consistent with the 2MASS magnitudes.

\section{Discussion}

Here, we discuss the implications of our results for all 20 {\em Chandra} 
observations of IGR source fields.  In the 12 cases where we find likely 
counterparts, these represent the first time that sub-arcsecond X-ray 
positions have been obtained.  We begin with discussions of the 6 cases 
where our {\em Chandra} positions confirm associations with previously 
suggested counterparts in the literature.  In these cases, counterparts 
were previously suggested based on the presence of unusual CVs, HMXBs, 
or AGN in the {\em INTEGRAL} error circles.  We then discuss the 6 cases 
where the {\em Chandra} detections either provide new optical/IR 
counterparts or rule out previously suggested counterparts.  Finally, we 
discuss the 8 cases where we were not able to obtain clear {\em Chandra} 
counterparts to the IGR sources.

\subsection{Confirmations of Previously Suggested Associations}

{\bf IGR J00234+6141:} The position of CXOU J002257.6+614107 
is consistent with the position of the Cataclysmic Variable (CV) 
reported in \cite{hm06} and \cite{bikmaev06}.  It is also less than 
$6^{\prime\prime}$ from the position of the {\em ROSAT} source 
1RXS~J002258.3+614111, confirming this association.  It has been found 
that this source belongs to the Intermediate Polar CV class and 
contains a white dwarf with a spin period of 563.5~s and an orbital
period of 4.033~h \citep{bb07}.

{\bf IGR J06074+2205:} The position of CXOU J060726.6+220547 
is consistent with the position of the Be star reported in 
\cite{ht05} and \cite{masetti06iv}, confirming that this source 
is a (Be-type) High-Mass X-ray Binary.  In addition to the 
position, our study shows that there is material local to the 
compact object that contributes to the column density of $N_{\rm H} = 
(7.2^{+2.5}_{-2.0})\times 10^{22}$ cm$^{-2}$ that we measure in 
the X-ray.  The fact that there must be local absorption is 
especially clear since the source is bright in the optical 
($B = 12.7\pm 0.3$).  When the source was originally detected by 
{\em INTEGRAL} in 2003, IGR~J06074+2205 was reported to have a 
3--10 keV flux of $7\pm 2$ mCrab \citep{chenevez04}, and we find 
that the flux during the {\em Chandra} observation was a factor 
of $\sim$50 lower (see Table~\ref{tab:comparison}).  Such a flux 
ratio between outburst and quiescence would not be unusual for a 
Be X-ray binary \citep{campana02}, and probably indicates a
binary orbit with some eccentricity.

{\bf IGR J10101--5654:} The position of CXOU J101011.8--565532 
is consistent with the position of the star that \cite{masetti06v} 
suggest as the possible counterpart.  \cite{masetti06v} report an 
optical spectrum for this star with a strong H$\alpha$ emission 
line and a reddened continuum, and they conclude that the star is
an early-type giant.  With a column density of $N_{\rm H} = 
(3.2^{+1.2}_{-1.0})\times 10^{22}$~cm$^{-2}$, the {\em Chandra} 
spectrum does not indicate a high level of local absorption.
It is notable that this is one of the sources for which the 
extrapolated {\em Chandra} flux is significantly lower than the 
flux measured by {\em INTEGRAL} (see Table~\ref{tab:comparison}).  
This indicates significant change in mass accretion rate onto the 
compact object, and it could point to an eccentric binary orbit.

{\bf IGR J11305--6256:} The position of CXOU J113106.9--625648
is consistent with the position of the star HD~100199, which was 
suggested as a possible counterpart by \cite{masetti06iii}.  As 
mentioned above, this source was the brightest companion we found 
with optical and IR magnitudes of $\sim$8.  The spectral type of 
the star is B0IIIe, so this system is a Be type HMXB, and 
\cite{masetti06iii} estimate its distance at $\sim$3 kpc.  The
{\em Chandra} energy spectrum shows a low level of absorption, 
$N_{\rm H} = (3.2^{+2.8}_{-2.2})\times 10^{21}$ cm$^{-2}$, consistent
with the low optical extinction and a relatively small distance
for the source.  At 3~kpc, the flux reported in Table~\ref{tab:spectra}
corresponds to a 0.3--10 keV luminosity of $(4.7^{+2.1}_{-3.6})\times
10^{34}$ ergs~s$^{-1}$.

{\bf IGR J14515--5542:} The position of CXOU J145133.1--554038
is consistent with the position of the Seyfert 2 Active Galactic
Nucleus LEDA 3079667, which has a redshift of $z = 0.018$
\citep{masetti06v}.  This source appears in the 2MASS catalog of
extended sources as 2MASX J14513316--5540388.  This source was
previously observed by {\em Swift}, and \cite{malizia07} mention
that this is one of five Seyfert 2 AGN in their sample with lower
levels of absorption than might be expected for Seyfert 2s.  
We measure a slightly higher value for $N_{\rm H}$ for our 
{\em Chandra} spectrum ($(1.0^{+0.4}_{-0.3})\times 10^{22}$ cm$^{-2}$
for {\em Chandra} vs. $(0.39^{+0.18}_{-0.16})\times 10^{22}$ cm$^{-2}$
for {\em Swift}).  Given that the value of $N_{\rm H}$ along this
line-of-sight through the Galaxy is $5.3\times 10^{21}$ cm$^{-2}$,
our result still confirms that the column density for this source 
is lower than is typically seen for Seyfert 2s.  

{\bf IGR J17200--3116:} The position of CXOU J172005.9--311659
is consistent with the position of the star that \cite{masetti06v}
suggest as a possible counterpart and is also consistent with the 
position of the {\em ROSAT} source 1RXS~J172006.1--311702, 
confirming this association.  \cite{masetti06v} report an 
optical spectrum for this star with an H$\alpha$ emission line 
and a reddened continuum, and they conclude that the system is a 
likely HMXB.  The {\em Chandra} energy spectrum indicates a
column density of $N_{\rm H} = (1.9^{+0.6}_{-0.5})\times 10^{22}$
cm$^{-2}$.  Although this is somewhat higher than the column 
density through the Galaxy along this line-of-sight, it does not
necessarily require local absorption.  

\subsection{IGR J09026--4812}

For this source, the {\em Chandra} observation described above in 
this paper led to the first X-ray position accurate enough to
obtain an optical/IR identification \citep{tomsick07_atel1}.
We have identified CXOU J090237.3--481334 with the 2MASS and
DENIS sources listed in Table~\ref{tab:oir}.  Although optical
or IR spectra are required to confirm that it is a Galactic 
object, we do not see any evidence that the source is extended
as might be expected for an AGN as IR-bright as this source
(e.g., IGR~J14515--5542).  Assuming that it is a Galactic source, 
it is probably relatively distant or highly reddened since it has 
the largest $J$--$K_{\rm s}$ value of any of our targets as mentioned 
above, and it is not detected in the USNO-B1.0 catalog or in the 
$I$-band in the DENIS catalog.  The $J$--$K_{\rm s}$ of $2.88\pm 0.09$
does not fit any stellar spectral type, and must be primarily due
to extinction.

If we hypothesize that the source is an HMXB, which may be required
for the combination of high extinction and IR brightness, we can 
perform a consistency check on our hypothesis by assuming
a spectral type of B0V, which has an absolute magnitude of
$M_{K} = -2.5$.  We assume that the extinction is at a level
consistent with the range between the Galactic value and the
X-ray value we measure (see Table~\ref{tab:spectra}), and this
gives $A_{V} = 5.8$--10.6 ($A_{K} = 0.66$--1.21).  Using 
$K_{\rm s} = 12.69$, we calculate a distance range of $d = 
6.3$--8.1~kpc, corresponding to a 0.3--10 keV X-ray luminosity 
of $(8^{+13}_{-1})\times 10^{34}$ ($d$/7~kpc)$^{2}$ ergs~s$^{-1}$, 
which is in-line with values obtained for other IGR HMXBs (and 
also the value of $\Gamma = 1.1^{+0.5}_{-0.3}$ for the X-ray 
spectral index is typical).  Overall, the hypothesis of an HMXB
with a main sequence companion works for IGR J09026--4812.  If
we instead assume a supergiant companion with $M_{K} = -5.9$
(B0I), we obtain a distance of 30--40~kpc, indicating that a
supergiant companion is unlikely.

\subsection{IGR J11435--6109}

IGR J11435--6109 is already known to be a neutron star HMXB based
on the detection of pulsations at 161.76~s, and a long-term X-ray
modulation at the orbital period of 52.36~d \citep{ih04,sm04,wen06}.
For this source, the {\em Chandra} observation described above in 
this paper led to the first X-ray position accurate enough to
obtain an optical/IR identification \citep{tomsick07_atel3}.
We have identified CXOU J114400.3--610736  with the 2MASS, 
DENIS, and USNO sources listed in Table~\ref{tab:oir}.  Soon
after we reported the {\em Chandra} position, a report of optical 
spectroscopy of the candidate was issued, suggesting a Be
companion for this HMXB \citep{negueruela07}.  The {\em Chandra}
spectrum indicates strong absorption with $N_{\rm H} = 
(1.5^{+0.5}_{-0.4})\times 10^{23}$ cm$^{-2}$, possibly due
to a strong stellar wind from the Be star.  Finally, although 
\cite{negueruela07} do not determine the stellar subclass, they 
calculate that if the companion is a B2Ve star, then its distance 
is $d > 6$~kpc.  Based on the {\em Chandra} spectrum, this corresponds 
to a lower limit on the 0.3--10~keV luminosity of $>$$3\times 10^{34}$ 
ergs~s$^{-1}$.

\subsection{IGR J18214--1318}

For this source, the {\em Chandra} observation described above in 
this paper led to the first X-ray position accurate enough to
obtain an optical/IR identification, and we have identified 
CXOU J182119.7--131838 with the 2MASS, DENIS, and USNO sources 
listed in Table~\ref{tab:oir}.  As shown in Figure~\ref{fig:images_2mass}i,
the source is blended with several other stars, and no strong
argument can be made for it being point-like or extended, so
we cannot rule out the possibility that it is an AGN.  If the 
source is Galactic, the relatively large $J$--$K_{s}$ color value 
of $2.4\pm 0.1$ must be caused by extinction since, as for 
IGR~J09026--4812, such a large value does not match any spectral
type.  This may imply a fairly large distance, but it is notable
that, unlike IGR~J09026--4812, IGR~J18214--1318 is detected in
the optical, and IGR~J18214--1318 is also somewhat brighter in 
the IR.  Although it is unclear whether this source is an HMXB, 
the X-ray spectrum does show evidence for local absorption with
a column density of $N_{\rm H} = (1.17^{+0.30}_{-0.27})\times 10^{23}$
cm$^{-2}$, and it would not be surprising if this was due to the
wind from a high-mass star.  Overall, we believe that the most
likely source type in this case is an HMXB, but spectral confirmation
is necessary.

\subsection{IGR J18256--1035}

Previously, \cite{landi07_atel} reported a {\em Swift} position for 
this source with a $4^{\prime\prime}.1$ error circle, and suggested an 
association between IGR~J18256--1035 and the optical source
USNO-B1.0~0794--399102.  While our {\em Chandra} position agrees
with the {\em Swift} position, we do not confirm the association 
with the USNO source as CXOU J182543.8--103501 is $2^{\prime\prime}.9$
away from the USNO source.  The image shown in Figure~\ref{fig:images_2mass}j
indicates that the region is quite crowded, and while the {\em Chandra}
source does not fall on any of the brighter $J$-band sources, it does
fall in a region where there is IR emission.  While IGR~J18256--1035
must be fainter than the nearest source in the 2MASS catalog, which has
$J\sim 15$, it may not be very much fainter than this.  Considering
the {\em Chandra} spectrum, this source is notable for having the
hardest (or one of the hardest when error bars are considered) power-law
indexes at $\Gamma = 0.1^{+0.5}_{-0.4}$.  However, the source is not
strongly absorbed, and its nature is not clear.  

\subsection{IGR J18259--0706}

This source has a reported {\em Swift} position with an error circle 
of $3^{\prime\prime}.71$ \citep{malizia07}.  In fact, the position of 
CXOU J182557.5--071022 is $5^{\prime\prime}.1$ away from the {\em Swift}
position even though it is still very likely that they are the same source.
Although the {\em Swift} position allows for the possibility that the bright 
IR source that is to the West of the {\em Chandra} source as shown in 
Figure~\ref{fig:images_2mass}k is the counterpart, the {\em Chandra} 
position indicates that the IGR source is the fainter source next to it.  
Follow-up optical spectroscopy was performed on this source based on 
the identification facilitated by our {\em Chandra} position.  
\cite{burenin08} report the presence of a broad and redshifted H$\alpha$
emission line in the optical spectrum, indicating that the source is an AGN.
Furthermore, \cite{burenin08} classify the source as a Seyfert 1.

\subsection{IGR J18325--0756}

Recently, \cite{landi08_atel} reported a {\em Swift} position for this 
source with a $4^{\prime\prime}.1$ error circle, and CXOU J183228.3--075641 
has a position that is consistent.  The 2MASS source that \cite{landi08_atel}
mention as a possible counterpart is the same source that we list in
Table~\ref{tab:oir}.  Although the {\em Chandra}/2MASS position difference
of $0^{\prime\prime}.625$ is somewhat larger than for our other 
identifications, our NTT observations appear to confirm the association 
(see Figure~\ref{fig:image_j18325}).  If the source is Galactic, the
relatively large $J$--$K_{s}$ color value of $2.0\pm 0.1$ is likely due 
to extinction and may suggest a relatively large distance.  The {\em Swift} 
spectrum that \cite{landi08_atel} describe is similar to the highly absorbed
spectrum we see as \cite{landi08_atel} mention the possibility of a column 
density as high as $1.5\times 10^{23}$ cm$^{-2}$.  Between our {\em Chandra} 
spectrum and the {\em Swift} spectrum, this provides evidence for local 
absorption in this source.  An HMXB nature should be considered as a 
possibility in this case.

\subsection{Discussion of Sources without Clear {\em Chandra} Counterparts}

{\bf IGR J01363+6610:}  We detected 3 {\em Chandra} sources within the 
{\em INTEGRAL} error circle for IGR~J01363+6610.  All three sources are very 
faint, with count rates between 0.0010 and 0.0019~c/s (see Table~\ref{tab:list}), 
and it is not possible to tell whether any of them are the correct counterpart
to the IGR source.  This IGR source has been suggested to be an HMXB with
a Be-type companion based on the fact that there is a Be star in the {\em INTEGRAL}
error circle \citep{reig05}.  While our {\em Chandra} field-of-view covers
the location of the Be star, we do not detect any X-ray source at this position.
This source is a documented transient as \cite{grebenev04} describe an {\em INTEGRAL}
detection at 17 mCrab and then a non-detection with an upper limit of 11 mCrab.
The source may not be extremely highly absorbed because it was detected by 
JEM-X during its 2004 outburst \citep{grebenev04}; however, the JEM-X detection
was reported in the 8--15~keV band, which still allows for the possibility of
significant absorption.  Based on the brightest of the 3 possible {\em Chandra}
counterparts (with a count rate of 0.0019~c/s), we have calculated an upper limit
to the source flux during our {\em Chandra} observation.  Assuming that $N_{\rm H}$
is at the Galactic value (see Table~\ref{tab:spectra}) and a power-law spectral
shape with a photon index between $\Gamma = 1$ and 2, we calculate an upper limit
on the unabsorbed 0.3--10~keV flux in the range (3.2--4.1)$\times 10^{-14}$ 
ergs~cm$^{-2}$~s$^{-1}$, which corrresponds to a luminosity of $\sim$$5\times 10^{30}$
$d_{\rm kpc}^{2}$ ergs~s$^{-1}$, where $d_{\rm kpc}$ is the source distance in kpc.
At the 2~kpc distance suggested by \cite{reig05} for the Be star, this would 
correspond to a luminosity of $2\times 10^{31}$ ergs~s$^{-1}$, which is significantly
lower than quiescent luminosities of other Be X-ray binaries \citep{campana02}.

{\bf IGR J17285--2922:} We detected 6 {\em Chandra} sources within the {\em INTEGRAL} 
error circle for IGR~J17285--2922 with a range of count rates between 0.0013 and 
0.0030~c/s (see Table~\ref{tab:list}).  It is not possible to tell whether any of them 
are the correct counterpart to the IGR source, and no optical counterpart has been
suggested for this source.  {\em INTEGRAL} observations of the source clearly show
that the source is a transient, and when it had an outburst in 2003, its power-law 
spectrum with a photon index of $\Gamma\sim 2.1$ suggest the possibility that it is 
a black hole transient.  To calculate a flux upper limit from our {\em Chandra} 
observation, we make the same assumptions as for the previous source, and find
that the upper limit on the unabsorbed 0.3--10~keV flux is in the range
(5.5--6.4)$\times 10^{-14}$ ergs~cm$^{-2}$~s$^{-1}$.  Such a low flux would not
be surprising for a black hole transient.

{\bf IGR J17331--2406:} Little is known about this source other than that it a
transient that had a $\sim$9 mCrab outburst in 2004 with a hard spectrum that was
fitted with a power-law with $\Gamma\sim 1.8$ \citep{lutovinov04}.  No optical 
counterpart has been suggested for this source, and we did not detect any 
{\em Chandra} sources in our observation.  This indicates a count rate upper 
limit of $\sim$0.0008~c/s, which corresponds to an upper limit on the unabsorbed
0.3--10~keV flux of (1.0--1.5)$\times 10^{-14}$ ergs~cm$^{-2}$~s$^{-1}$.  As for 
the previous source, such a flux level could be consistent with a black hole
transient nature, but we cannot rule out other possibilities.

{\bf IGR J17407--2808:}  We detected 7 {\em Chandra} sources within the {\em INTEGRAL} 
error circle for IGR~J17407--2808 with a range of count rates between 0.0011 and 
0.0031~c/s.  It is not possible to tell whether any of them are the correct counterpart 
to the IGR source, and no optical counterpart has been suggested for this source.
The possible X-ray counterpart 2RXP~J174040.9--280852 was suggested as a possible 
counterpart, but we do not detect a {\em Chandra} source within the $16^{\prime\prime}$ 
error circle.  The source is known to be transient, and based on {\em INTEGRAL} 
detections of short flares as bright as 800 mCrab, it was suggested that the source 
is a Supergiant Fast X-ray Transient \citep{sguera06}.  The ACIS-S count rate upper 
limit of 0.0031~c/s from our {\em Chandra} observation corresponds to an upper limit 
on the absorbed flux of (6.0--7.2)$\times 10^{-14}$ ergs~cm$^{-2}$~s$^{-1}$.  Some 
other SFXTs have been shown to be quite faint in quiescence, with X-ray luminosities 
as low as $5\times 10^{32}$ ergs~s$^{-1}$ \citep{intzand05}, so the low {\em Chandra}
flux for IGR J17407--2808 may be consistent with a SFXT nature.  However, in this
case, it should also be noted that the {\em INTEGRAL} error circle is relatively 
large for this IGR source, with a radius of $4^{\prime}.2$.  While the {\em Chandra}
field-of-view does fully cover the error circle, if the IGR source actually lies
slightly outside of the error circle, it could have been missed.

{\bf IGR J17445--2747:}  We detected 2 {\em Chandra} sources within the {\em INTEGRAL} 
error circle for this source with count rates of 0.0008 and 0.0017~c/s, and it is not 
possible to tell whether either of these is the correct counterpart to the IGR source.
In fact, very little is known about this source.  While it might be a transient, its
flux history has not been well-documented.  The suggested association with a {\em Swift}
source is discussed above in \S~2.2, and while we detect the same source with
{\em Chandra}, the fact that both of these sources are well outside the {\em INTEGRAL}
error circle makes the connection dubious.  Assuming that the {\em Swift} source is
not the correct counterpart, we calculate an upper limit on the 0.3--10~keV
unabsorbed flux of (4.1--4.5)$\times 10^{-14}$ ergs~cm$^{-2}$~s$^{-1}$.

{\bf IGR J17507--2856:} We detected 3 {\em Chandra} sources within the {\em INTEGRAL} 
error circle for this source.  All three sources are faint, with count rates between 
0.0012 and 0.0033~c/s, and it is not possible to tell whether any of them are the 
correct counterpart to the IGR source.  The source is called a transient by
\cite{gs04}, but its flux history has not been well-documented.  Also, there have 
not been any suggestions of optical counterparts for this source.  We derive an
upper limit on the unabsorbed 0.3--10~keV flux of (7.4--8.4)$\times 10^{-14}$ 
ergs~cm$^{-2}$~s$^{-1}$.

{\bf IGR J18193--2542:}  We did not detect any {\em Chandra} source in the {\em INTEGRAL}
error circle for IGR~J18193--2542, and its nature is unclear.  The source was listed
in the second IBIS catalog \citep{bird06}, but not the third IBIS catalog 
\citep{bird07}.  It is listed as a transient in \citep{bodaghee07}, but its flux
history is not well-known.  No optical counterpart for the source has been detected.  
Based on a non-detection in our {\em Chandra} observation, we infer an upper limit on 
the 0.3--10~keV unabsorbed flux of between $8.9\times 10^{-15}$ and 
$1.4\times 10^{-14}$ ergs~cm$^{-2}$~s$^{-1}$.

{\bf IGR J18539+0727:} There is good evidence that this source is a black hole transient
as \cite{lr03} were able to use the {\em Rossi X-ray Timing Explorer (RXTE)} to measure
its spectral and timing properties during its outburst.  Thus, perhaps it is not too
surprising that we did not detect any source in the {\em INTEGRAL} error circle with
{\em Chandra} because black hole transients tend to be very faint when they are in
quiescence.  Based on a non-detection in our {\em Chandra} observation, we infer
an upper limit on the 0.3--10~keV unabsorbed flux in the range (1.7--2.0)$\times 10^{-14}$
ergs~cm$^{-2}$~s$^{-1}$, which corresponds to an X-ray luminosity limit of 
$<$$2\times 10^{32}$ ergs~s$^{-1}$ if the source distance is 10~kpc.  

\section{Conclusions}

During {\em Chandra} observing cycles 6 and 8, we have obtained snapshot observations
of 24 IGR sources located within $5^{\circ}$ of the Galactic plane.  In 16 cases, a
{\em Chandra} counterpart is detected, providing the first sub-arcsecond X-ray positions 
for these sources.  Such an accurate position is required to obtain or confirm optical 
or IR counterparts in the crowded Galactic plane region, and a summary of the results 
for these 16 sources is given in Table~\ref{tab:summary}.  The information in the table
includes the position in Galactic co-ordinates, the $J$-band magnitude, the source 
type, the spectral type of the optical companion for Galactic sources or the AGN type
for the one AGN in our sample, and an estimate of the contribution to the column density
that is local to the system.  For this last quantity, $N_{\rm H,local}$, we derive a
range of possible levels of local absorption from the values given in 
Table~\ref{tab:spectra} and from \cite{tomsick06}, where the upper limit comes from
assuming that all of the measured $N_{\rm H}$ is local and the lower limit comes
from subtracting the Galactic $N_{\rm H}$ and two times the Galactic $N_{\rm H_{2}}$
from the measured value.  If the lower limit is less than or consistent with zero, 
then there is no evidence for local absorption, and only the upper limit is quoted.
For Galactic sources, a high level of local absorption provides evidence that a
strong stellar wind is enveloping the compact object.

Considering the different types of sources with {\em Chandra} counterparts in our
sample, the HMXBs are the largest group, with 6 of the sources being spectroscopically 
confirmed HMXBs, and 4 of the sources being possible HMXBs.  At least 3 of the 
confirmed HMXBs are Be X-ray binaries (IGR~J06074+2205, IGR~J11435--6109, and 
IGR~J11305--6256), and the {\em Chandra} spectra for the first two show evidence for 
significant local absorption ($N_{\rm H,local}$ is in the range (4.6--9.7)$\times 
10^{22}$ and (10--20)$\times 10^{22}$~cm$^{-2}$ for IGR~J06074+2205 and IGR~J11435--6109, 
respectively).  IGR~J16207--5129 has an early-type (i.e., high-mass) supergiant 
companion with no evidence for local aborption, and IGR~J10101--5645 and IGR~J17200--3116 
have early-type companions, but better optical or IR spectra are required to determine
the exact spectral type.   

Our determination of which of the remaining sources are possible or likely HMXBs depend 
on several measurements that we report in this paper.  All 4 of these systems have 
hard X-ray spectra and none of the IR images indicate that they are extended sources.
In addition, IGR~J09026--4812 is bright in the IR ($K_{\rm s}\sim 13$), but it shows 
relatively high optical/IR extinction, suggesting that it is relatively distant with 
a luminous companion.  A similar argument based on the optical/IR magnitudes holds 
for IGR~J18214--1318, IGR~J18325--0756, and IGR~J16195--4945.  In addition, the 
evidence that IGR~J18214--1318 and IGR~J18325--0756 are HMXBs is stronger because of
the fact that $N_{\rm H,local}$ is in the range (7--15)$\times 10^{22}$~cm$^{-2}$ 
for the former and in the range (6--78)$\times 10^{22}$~cm$^{-2}$ for the latter, 
possibly indicating a strong stellar wind.  

Of the remaining 6 sources, 3 of them (IGR~J00234+6141, IGR~J16167--4957, and 
IGR~J17195--4100) are spectroscopically confirmed CVs.  All 3 of these sources have
very low X-ray column densities and are likely to be relatively nearby.  
IGR~J14515--5542 and IGR~J18259--0706 are AGN.  While IGR~J14515--5542 has been 
spectroscopically classified as a Seyfert 2, our {\em Chandra} spectrum confirms an 
earlier {\em Swift} result that the X-ray column density is lower than would be expected 
from a Seyfert 2.  The final source with a {\em Chandra} counterpart is IGR~J18256--1035.  
Our results confirm that this source has a hard X-ray spectrum, but it does not 
necessarily have any local absorption.  Also, it is in a region that is very crowded 
with IR sources, and better optical/IR images are necessary to determine its brightness.

Table~\ref{tab:summary2} lists the remaining 8 sources without a clear {\em Chandra}
counterpart along with the upper limits on the X-ray flux from {\em Chandra} as
well as any information about the source type from the literature.  In 3 cases
(IGR~J17285--2922, IGR~J17331--2406, and IGR~J18539+0727), there is evidence
from earlier {\em INTEGRAL} observations that these sources are black hole candidate
(BHC) X-ray transients.  Such systems can be very faint in quiescence 
\citep{garcia01,tomsick03_qbh}, so our {\em Chandra} non-detections can be seen
as further evidence for their BHC nature.  The non-detections for the likely HMXBs, 
IGR~J01363+6610 and IGR J17407--2808, are more surprising.  While many HMXBs show 
high levels of variability, flux upper limits in the range of $10^{-13}$ to 
$10^{-14}$ ergs~cm$^{-2}$~s$^{-1}$ are lower than expected, and may indicate
that these systems are more distant or have atypical properties.  Although we
do not know the source type for the final 3 sources (IGR~J17445--2747, 
IGR~J17507--2856, and IGR~J18193--2542), we note that all 3 sources have locations
in the vicinity of the Galactic center.  This could indicate that they are
part of bulge population of Low-Mass X-ray Binary (possibly BHC) transients.

\acknowledgments

JAT acknowledges partial support from {\em Chandra} award number
GO7-8050X issued by the {\em Chandra X-Ray Observatory Center}, which
is operated by the Smithsonian Astrophysical Observatory for and on 
behalf of the National Aeronautics and Space Administration (NASA), 
under contract NAS8-03060.  JAT acknowledges partial support from a 
NASA {\em INTEGRAL} Guest Observer grant NNX07AQ13G.  This work is
based in part on observations collected at the European Organization
for Astronomical Research in the Southern Hemisphere, Chile (ESO
program 073.D-0339).  This publication makes use of data products 
from the Two Micron All Sky Survey, which is a joint project of the 
University of Massachusetts and the Infrared Processing and Analysis 
Center/California Institute of Technology, funded by NASA and the 
National Science Foundation.  This research makes use of the USNOFS 
Image and Catalogue Archive operated by the United States Naval 
Observatory, Flagstaff Station, the SIMBAD database, operated at CDS, 
Strasbourg, France, and the Deep Near Infrared Survey of the Southern 
Sky (DENIS).



\clearpage

\begin{table}
\caption{{\em Chandra} Observations\label{tab:obs}}
\begin{minipage}{\linewidth}
\footnotesize
\begin{tabular}{cccclc} \hline \hline
IGR Name & ObsID & $l$\footnote{Galactic longitude in degrees.} & $b$\footnote{Galactic latitude in degrees.} & Start Time & Exposure Time (s)\\ \hline \hline
J00234+6141  & 7524 & 119.62 & --1.00 & 2007 Mar 26, 3.18 h UT   & 4891\\
J01363+6610  & 7533 & 127.45 &  +3.70 & 2007 Jun 8, 4.16 h UT    & 4976\\
J06074+2205  & 7520 & 188.39 &  +0.80 & 2006 Dec 2, 12.52 h UT   & 4935\\
J09026--4812 & 7525 & 268.88 & --1.09 & 2007 Feb 5, 0.36 h UT    & 4692\\
J10101--5645 & 7519 & 282.24 & --0.67 & 2007 Jun 9, 9.98 h UT    & 4698\\
J11305--6256 & 7527 & 293.87 & --1.49 & 2007 Sept 27, 2.22 h UT  & 5058\\
J11435--6109 & 7523 & 296.05 &  +0.97 & 2007 Sept 23, 10.14 h UT & 5093\\
J14515--5542 & 7531 & 319.34 &  +3.29 & 2007 May 16, 7.23 h UT   & 4888\\
J17200--3116 & 7532 & 355.02 &  +3.35 & 2007 Sept 30, 22.40 h UT & 4692\\
J17285--2922 & 7530 & 357.64 &  +2.88 & 2008 Feb 8, 19.45 h UT   & 4698\\
J17331--2406 & 7535 &   2.61 &  +4.93 & 2008 Feb 5, 6.72 h UT    & 5106\\
J17407--2808 & 7526 &   0.10 &  +1.34 & 2007 Aug 1, 13.08 h UT   & 5109\\
J17445--2747 & 7521 &   0.86 &  +0.81 & 2008 Feb 8, 17.57 h UT   & 4888\\
J17507--2856 & 7562 &   0.58 & --0.96 & 2008 Feb 5, 8.52 h UT    & 4695\\
J18193--2542 & 7534 &   6.51 & --4.91 & 2008 Feb 16, 3.89 h UT   & 5084\\
J18214--1318 & 7561 &  17.73 &  +0.47 & 2008 Feb 14, 13.06 h UT  & 4692\\
J18256--1035 & 7522 &  20.58 &  +0.84 & 2008 Feb 14, 11.48 h UT  & 4698\\
J18259--0706 & 7528 &  23.64 &  +2.35 & 2008 Feb 14, 9.70 h UT   & 4698\\ 
J18325--0756 & 7518 &  23.72 &  +0.56 & 2008 Feb 26, 22.79 h UT  & 5061\\ 
J18539+0727  & 7529 &  39.85 &  +2.85 & 2007 Feb 19, 11.23 h UT  & 4656\\ \hline
\end{tabular}
\end{minipage}
\end{table}

\begin{table}
\caption{{\em Chandra} Identifications\label{tab:sources}}
\begin{minipage}{\linewidth}
\footnotesize
\begin{tabular}{ccccc} \hline \hline
IGR Name\footnote{For all sources, the largest contribution to the uncertainties in the {\em Chandra} positions are due to pointing systematics.  The pointing uncertainties are $0^{\prime\prime}\!.64$ at 90\% confidence and $1^{\prime\prime}$ at 99\% confidence \citep{weisskopf05}.  The statistical position uncertainties are between $0^{\prime\prime}\!.02$ and $0^{\prime\prime}\!.11$.} & {\em Chandra} & {\em Chandra} & {\em Chandra}/ACIS\\
         & R.A. (J2000)  & Decl. (J2000) &  Count Rate\\ \hline
J00234+6141  & $00^{\rm h}22^{\rm m}57^{\rm s}\!.64$ &  +$61^{\circ}41^{\prime}07^{\prime\prime}\!.5$ & 0.446\\
J06074+2205  & $06^{\rm h}07^{\rm m}26^{\rm s}\!.62$ &  +$22^{\circ}05^{\prime}47^{\prime\prime}\!.6$ & 0.064\\
J09026--4812 & $09^{\rm h}02^{\rm m}37^{\rm s}\!.33$ & --$48^{\circ}13^{\prime}34^{\prime\prime}\!.1$ & 0.235\\
J10101--5645 & $10^{\rm h}10^{\rm m}11^{\rm s}\!.87$ & --$56^{\circ}55^{\prime}32^{\prime\prime}\!.1$ & 0.055\\
J11305--6256 & $11^{\rm h}31^{\rm m}06^{\rm s}\!.95$ & --$62^{\circ}56^{\prime}48^{\prime\prime}\!.9$ & 0.203\\
J11435--6109 & $11^{\rm h}44^{\rm m}00^{\rm s}\!.31$ & --$61^{\circ}07^{\prime}36^{\prime\prime}\!.5$ & 0.074\\
J14515--5542 & $14^{\rm h}51^{\rm m}33^{\rm s}\!.15$ & --$55^{\circ}40^{\prime}38^{\prime\prime}\!.4$ & 0.223\\
J17200--3116 & $17^{\rm h}20^{\rm m}05^{\rm s}\!.92$ & --$31^{\circ}16^{\prime}59^{\prime\prime}\!.4$ & 0.215\\
J18214--1318 & $18^{\rm h}21^{\rm m}19^{\rm s}\!.76$ & --$13^{\circ}18^{\prime}38^{\prime\prime}\!.9$ & 0.197\\
J18256--1035 & $18^{\rm h}25^{\rm m}43^{\rm s}\!.83$ & --$10^{\circ}35^{\prime}01^{\prime\prime}\!.9$ & 0.057\\

J18259--0706 & $18^{\rm h}25^{\rm m}57^{\rm s}\!.58$ & --$07^{\circ}10^{\prime}22^{\prime\prime}\!.8$ & 0.487\\ 

J18325--0756 & $18^{\rm h}32^{\rm m}28^{\rm s}\!.32$ & --$07^{\circ}56^{\prime}41^{\prime\prime}\!.7$ & 0.0059\\ \hline
\end{tabular}
\end{minipage}
\end{table}

\begin{table}
\caption{{\em Chandra} source list for cases without a clear counterpart\label{tab:list}}
\begin{minipage}{\linewidth}
\footnotesize
\begin{tabular}{cccc} \hline \hline
CXOU Name\footnote{For all sources, the position errors given in the table are the 1-$\sigma$ statistical errors calculated by the {\tt wavdetect} software.  A systematic pointing uncertainty should be added to this.  The pointing uncertainties are $0^{\prime\prime}\!.64$ at 90\% confidence and $1^{\prime\prime}$ at 99\% confidence \citep{weisskopf05}.} & {\em Chandra} R.A. & {\em Chandra} Decl. & {\em Chandra}/ACIS\\
         & (J2000)  & (J2000) &  Count Rate\\ \hline
\multicolumn{4}{c}{IGR J01363+6610}\\ \hline
J013609.9+661157 &  $01^{\rm h}36^{\rm m}09^{\rm s}.99\pm 0^{\rm s}.03$ & +$66^{\circ}11^{\prime}57^{\prime\prime}\!.5\pm 0^{\prime\prime}\!.2$ & 0.0010\\
J013621.2+660928 &  $01^{\rm h}36^{\rm m}21^{\rm s}.22\pm 0^{\rm s}.06$ & +$66^{\circ}09^{\prime}28^{\prime\prime}\!.7\pm 0^{\prime\prime}\!.2$ & 0.0012\\
J013632.8+660924 &  $01^{\rm h}36^{\rm m}32^{\rm s}.84\pm 0^{\rm s}.07$ & +$66^{\circ}09^{\prime}24^{\prime\prime}\!.3\pm 0^{\prime\prime}\!.2$ & 0.0019\\ \hline
\multicolumn{4}{c}{IGR J17285--2922}\\ \hline
J172830.1--292334 & $17^{\rm h}28^{\rm m}30^{\rm s}.18\pm 0^{\rm s}.02$ & --$29^{\circ}23^{\prime}34^{\prime\prime}\!.7\pm 0^{\prime\prime}\!.2$ & 0.0019\\
J172830.5--291946 & $17^{\rm h}28^{\rm m}30^{\rm s}.59\pm 0^{\rm s}.01$ & --$29^{\circ}19^{\prime}46^{\prime\prime}\!.6\pm 0^{\prime\prime}\!.2$ & 0.0013\\
J172831.1--292250 & $17^{\rm h}28^{\rm m}31^{\rm s}.17\pm 0^{\rm s}.02$ & --$29^{\circ}22^{\prime}50^{\prime\prime}\!.9\pm 0^{\prime\prime}\!.2$ & 0.0023\\
J172834.2--292458 & $17^{\rm h}28^{\rm m}34^{\rm s}.22\pm 0^{\rm s}.01$ & --$29^{\circ}24^{\prime}58^{\prime\prime}\!.0\pm 0^{\prime\prime}\!.2$ & 0.0016\\
J172836.5--291926 & $17^{\rm h}28^{\rm m}36^{\rm s}.52\pm 0^{\rm s}.02$ & --$29^{\circ}19^{\prime}26^{\prime\prime}\!.2\pm 0^{\prime\prime}\!.2$ & 0.0018\\
J172837.8--292133 & $17^{\rm h}28^{\rm m}37^{\rm s}.80\pm 0^{\rm s}.01$ & --$29^{\circ}21^{\prime}33^{\prime\prime}\!.6\pm 0^{\prime\prime}\!.1$ & 0.0030\\ \hline
\multicolumn{4}{c}{IGR J17407--2808}\\ \hline
J174024.9--281243 & $17^{\rm h}40^{\rm m}24^{\rm s}.95\pm 0^{\rm s}.03$ & --$28^{\circ}12^{\prime}43^{\prime\prime}\!.5\pm 0^{\prime\prime}\!.3$ & 0.0023\\
J174031.5--281326 & $17^{\rm h}40^{\rm m}31^{\rm s}.58\pm 0^{\rm s}.02$ & --$28^{\circ}13^{\prime}26^{\prime\prime}\!.8\pm 0^{\prime\prime}\!.3$ & 0.0013\\
J174032.3--281042 & $17^{\rm h}40^{\rm m}32^{\rm s}.34\pm 0^{\rm s}.04$ & --$28^{\circ}10^{\prime}42^{\prime\prime}\!.0\pm 0^{\prime\prime}\!.2$ & 0.0031\\
J174036.4--280840 & $17^{\rm h}40^{\rm m}36^{\rm s}.41\pm 0^{\rm s}.04$ & --$28^{\circ}08^{\prime}40^{\prime\prime}\!.3\pm 0^{\prime\prime}\!.3$ & 0.0026\\
J174043.2--281601 & $17^{\rm h}40^{\rm m}43^{\rm s}.27\pm 0^{\rm s}.02$ & --$28^{\circ}16^{\prime}01^{\prime\prime}\!.3\pm 0^{\prime\prime}\!.3$ & 0.0026\\
J174044.7--281159 & $17^{\rm h}40^{\rm m}44^{\rm s}.74\pm 0^{\rm s}.02$ & --$28^{\circ}11^{\prime}59^{\prime\prime}\!.5\pm 0^{\prime\prime}\!.2$ & 0.0012\\
J174054.6--280949 & $17^{\rm h}40^{\rm m}54^{\rm s}.60\pm 0^{\rm s}.01$ & --$28^{\circ}09^{\prime}49^{\prime\prime}\!.5\pm 0^{\prime\prime}\!.5$ & 0.0011\\ \hline
\multicolumn{4}{c}{IGR J17445--2747}\\ \hline
J174427.3--274324 & $17^{\rm h}44^{\rm m}27^{\rm s}.33\pm 0^{\rm s}.03$ & --$27^{\circ}43^{\prime}24^{\prime\prime}\!.1\pm 0^{\prime\prime}\!.4$ & 0.0008\\
J174435.4--274453 & $17^{\rm h}44^{\rm m}35^{\rm s}.40\pm 0^{\rm s}.01$ & --$27^{\circ}44^{\prime}53^{\prime\prime}\!.3\pm 0^{\prime\prime}\!.2$ & 0.0017\\ \hline
\multicolumn{4}{c}{IGR J17507--2856}\\ \hline
J175036.7--285452 & $17^{\rm h}50^{\rm m}36^{\rm s}.76\pm 0^{\rm s}.02$ & --$28^{\circ}54^{\prime}52^{\prime\prime}\!.6\pm 0^{\prime\prime}\!.3$ & 0.0030\\
J175045.3--285817 & $17^{\rm h}50^{\rm m}45^{\rm s}.30\pm 0^{\rm s}.02$ & --$28^{\circ}58^{\prime}17^{\prime\prime}\!.0\pm 0^{\prime\prime}\!.4$ & 0.0012\\
J175049.7--285510 & $17^{\rm h}50^{\rm m}49^{\rm s}.74\pm 0^{\rm s}.02$ & --$28^{\circ}55^{\prime}10^{\prime\prime}\!.7\pm 0^{\prime\prime}\!.2$ & 0.0033\\ \hline
\end{tabular}
\end{minipage}
\end{table}

\begin{table}
\caption{{\em Chandra} Spectral Results\label{tab:spectra}}
\begin{minipage}{\linewidth}
\footnotesize
\begin{tabular}{cccccccc} \hline \hline
 & $N_{\rm H}$ &     & X-ray &  & PSF & Fit & Galactic $N_{\rm H}$/$N_{\rm H_{2}}$\\
IGR Name\footnote{The parameters are for power-law fits to the {\em Chandra}/ACIS spectra and include photoelectric absorption with \cite{wam00} abundances.  A pile-up correction was applied in the cases where pile-up parameters are given.  We performed fits without re-binning the data and using Cash statistics.  Errors in this table are at the 90\% confidence level ($\Delta$$C = 2.7$).} & ($\times 10^{22}$ cm$^{-2}$) & $\Gamma$ & Flux\footnote{Unabsorbed 0.3--10 keV flux in units of $10^{-12}$ erg~cm$^{-2}$~s$^{-1}$.} & $\alpha$\footnote{The grade migration parameter in the pile-up model \citep{davis01}.  The probability that $n$ events will be piled together but will still be retained after data filtering is $\alpha^{n-1}$.} &  Fraction\footnote{The fraction of the point spread function (PSF) treated for pile-up \citep[see][]{davis01}.  This parameter is required to be in the range 0.85--1.0.} & Statistic\footnote{The Cash statistic and degrees of freedom for the best fit model.} & ($\times 10^{22}$ cm$^{-2}$)\footnote{The atomic hydrogen column density through the Galaxy from \cite{dl90}.  We also give the molecular hydrogen column density through the Galaxy, using a CO map and conversion to $N_{\rm H_{2}}$ \citep{dht01}.}\\ \hline \hline
J00234+6141  & $0.17^{+0.05}_{-0.04}$ & $0.87\pm 0.09$      & $8.6\pm 0.5$           & -- & -- & 631.8/660 & 0.73/0.0009\\
J06074+2205  & $7.2^{+2.5}_{-2.0}$    & $0.9\pm 0.5$        & $3.5^{+0.8}_{-0.5}$    & -- & -- & 548.1/660 & 0.61/0.01\\
J09026--4812 & $1.9^{+0.6}_{-0.4}$    & $1.1^{+0.5}_{-0.3}$ & $13^{+23}_{-2}$         & $0.66^{+0.30}_{-0.24}$ & $0.85^{+0.03}_{-0.00}$ & 778.3/658 & 1.0/0.8\\
J10101--5645 & $3.2^{+1.2}_{-1.0}$ & $1.0^{+0.5}_{-0.4}$    & $2.03^{+0.35}_{-0.25}$ & -- & -- & 465.9/660 & 1.8/0.8\\
J11305--6256 & $0.32^{+0.28}_{-0.22}$ & $0.33^{+0.40}_{-0.28}$ & $44^{+20}_{-34}$    & $0.78^{+0.22}_{-0.20}$ & $0.91^{+0.09}_{-0.06}$ & 783.6/658 & 1.5/0.06\\
J11435--6109 & $15^{+5}_{-4}$         & $1.1\pm 0.6$        & $9.1^{+4.8}_{-1.7}$                   & -- & -- & 474.4/660 & 1.0/0.02\\
J14515--5542 & $1.0^{+0.4}_{-0.3}$ & $1.5^{+0.9}_{-0.6}$    & $14^{+15}_{-7}$        & $0.66^{+0.34}_{-0.14}$ & $0.91^{+0.07}_{-0.05}$ & 790.8/658 & 0.53/0.002\\
J17200--3116 & $1.9^{+0.6}_{-0.5}$ & $0.8^{+0.5}_{-0.4}$ & $26^{+39}_{-16}$ & $0.45^{+0.55}_{-0.10}$ & $0.87^{+0.03}_{-0.02}$ & 727.7/658 & 0.49/0.07\\
J18214--1318 & $11.7^{+3.0}_{-2.7}$ & $0.7^{+0.6}_{-0.5}$ & $60^{+317}_{-35}$ & $0.36^{+0.64}_{-0.10}$ & $0.87^{+0.04}_{-0.02}$ & 634.2/658 & 1.6/0.4\\
J18256--1035 & $3.1^{+1.6}_{-1.2}$ & $0.1^{+0.5}_{-0.4}$ & $2.9^{+0.5}_{-0.4}$                      & -- & -- & 503.5/660 & 1.4/0.2\\
J18259--0706 & $1.7\pm 0.2$        & $1.02\pm 0.13$      & $14.3\pm 0.6$                            & -- & -- & 682.9/660 & 0.71/0.4\\
J18325--0756 & $34^{+44}_{-23}$    & $2.8^{+5.4}_{-2.7}$ & $6.0^{+1.8\times 10^{7}}_{-5.5}$         & -- & -- & 155.4/660 & 1.8/1.7\\ \hline
\end{tabular}
\end{minipage}
\end{table}

\begin{table}
\caption{Comparison between {\em INTEGRAL} and {\em Chandra} Fluxes\label{tab:comparison}}
\begin{minipage}{\linewidth}
\footnotesize
\begin{tabular}{ccccc} \hline \hline
         & Marked as Transient & Flux measured & Flux Extrapolated & Flux\\
IGR Name & in the IBIS Catalog & by {\em INTEGRAL}\footnote{The 20--40 keV (or 3--10 keV in the case of IGR J06074+2205) {\em INTEGRAL} fluxes in ergs~cm$^{-2}$~s$^{-1}$ from the IBIS Catalog \citep{bird07}.} & from {\em Chandra} Spectra\footnote{These fluxes, in ergs~cm$^{-2}$~s$^{-1}$, are calculated by extrapolating the power-law fits to the {\em Chandra} spectra from Table~\ref{tab:spectra} into the 20--40 keV band (or into the 3--10 keV band for IGR J06074+2205).}  & Ratio\footnote{The ratio of the extrapolated flux from the {\em Chandra} spectra to the flux measured by {\em INTEGRAL}.}\\ \hline\hline
J00234+6141  & N & $(3.8\pm 0.8)\times 10^{-11}$ & $(2.3\pm 0.4)\times 10^{-11}$ & $0.61\pm 0.17$\\
J06074+2205\footnote{This source was detected as a transient by JEM-X \citep{chenevez04} in the 3--10 keV band.  It has not been detected by IBIS, and does not appear in the IBIS catalog.} & Y & $(1.3\pm 0.4)\times 10^{-10}$ & $(2.6\pm 0.7)\times 10^{-12}$ & $0.021\pm 0.009$\\
J09026--4812 & N & $(9.8\pm 0.8)\times 10^{-11}$ & $(2.2^{+4.3}_{-1.8})\times 10^{-11}$ & $0.22^{+0.44}_{-0.18}$\\
J10101--5645 & N & $(8.3\pm 0.8)\times 10^{-11}$ & $(4.2\pm 3.4)\times 10^{-12}$ & $0.05\pm 0.04$\\
J11305--6256 & N & $(3.0\pm 0.1)\times 10^{-10}$ & $(3.1\pm 2.4)\times 10^{-10}$ & $1.0\pm 0.8$\\
J11435--6109 & N & $(9.1\pm 1.5)\times 10^{-11}$ & $(1.5^{+2.0}_{-1.5})\times 10^{-11}$ & $0.17^{+0.23}_{-0.17}$\\
J14515--5542 & N & $(6.8\pm 0.8)\times 10^{-11}$ & $(9.9^{+20.3}_{-9.9})\times 10^{-12}$ & $0.15^{+0.30}_{-0.15}$\\
J17200--3116 & Y & $(2.1\pm 0.1)\times 10^{-10}$ & $(7.9^{+13.5}_{-7.9})\times 10^{-11}$ & $0.38^{+0.64}_{-0.38}$\\
J18214--1318 & Y & $(1.3\pm 0.1)\times 10^{-10}$ & $(2.2^{+11.8}_{-2.2})\times 10^{-10}$ & $1.7^{+5.4}_{-1.7}$\\
J18256--1035 & N & $(7.6\pm 0.8)\times 10^{-11}$ & $(3.0\pm 2.2)\times 10^{-11}$ & $0.39\pm 0.29$\\
J18259--0706 & N & $(7.6\pm 0.8)\times 10^{-11}$ & $(2.8\pm 0.7)\times 10^{-11}$ & $0.37\pm 0.10$\\
J18325--0756 & Y & $(1.9\pm 0.1)\times 10^{-10}$ & $<$$5\times 10^{-12}$\footnote{In this case, the large errors on the {\em Chandra} power-law parameters make a robust extrapolation into the 20--40 keV band impossible.  This upper limit corresponds to the case of $\Gamma = 0.1$, which is the hardest power-law index consistent with the {\em Chandra} spectrum.}  & $<$0.03\\ \hline
\end{tabular}
\end{minipage}
\end{table}

\begin{table}
\caption{Optical/Infrared Identifications\label{tab:oir}}
\begin{minipage}{\linewidth}
\footnotesize
\begin{tabular}{ccccc} \hline \hline
Catalog/Source\footnote{The catalogs are the 2 Micron All-Sky Survey (2MASS), 
the Deep Near Infrared Survey of the Southern Sky (DENIS), and
the United States Naval Observatory (USNO-B1.0).  For IGR~J18256--1035
we did not find a good 2MASS/{\em Chandra} association, and we listed
the closest 2MASS source.} & Separation & & Magnitudes & \\ \hline
\multicolumn{5}{c}{IGR J00234+6141/CXOU J002257.6+614107}\\ \hline
2MASS J00225764+6141075 & $0^{\prime\prime}.090$ & $J = 15.122\pm 0.048$ & $H = 15.046\pm 0.082$ & $K_{s} = 14.770\pm 0.124$\\
USNO-B1.0 1516--0012283 & $0^{\prime\prime}.325$ & $B = 18.0\pm 0.3$ & $R = 16.3\pm 0.3$ & $I = 15.6\pm 0.3$\\ \hline
\multicolumn{5}{c}{IGR J06074+2205/CXOU J060726.6+220547}\\ \hline
2MASS J06072661+2205477 & $0^{\prime\prime}.227$ & $J = 10.491\pm 0.021$ & $H = 10.189\pm 0.022$ & $K_{s} = 9.961\pm 0.019$\\
USNO-B1.0 1120--0112757 & $0^{\prime\prime}.467$ & $B = 12.7\pm 0.3$ & $R = 11.3\pm 0.3$ & $I = 10.2\pm 0.3$\\ \hline
\multicolumn{5}{c}{IGR J09026--4812/CXOU J090237.3--481334}\\ \hline
2MASS J09023731--4813339 & $0^{\prime\prime}.181$ & $J = 15.568\pm 0.081$ & $H = 13.863\pm 0.071$ & $K_{s} = 12.687\pm 0.040$\\
DENIS J090237.3--481334  & $0^{\prime\prime}.451$ & ---                   & $J = 15.831\pm 0.17$  & $K_{s} = 13.011\pm 0.15$\\ \hline
\multicolumn{5}{c}{IGR J10101--5654/CXOU J101011.8--565532}\\ \hline
2MASS J10101186--5655320 & $0^{\prime\prime}.006$ & $J = 12.617\pm 0.044$ & $H = 11.480\pm 0.041$ & $K_{s} = 10.669\pm 0.029$\\
DENIS J101011.8--565532  & $0^{\prime\prime}.413$ & $I = 15.632\pm 0.06$ &  $J = 12.506\pm 0.07$  & $K_{s} = 10.657\pm 0.07$\\ \hline
\multicolumn{5}{c}{IGR J11305--6256/CXOU J113106.9--625648}\\ \hline
2MASS J11310691--6256489 & $0^{\prime\prime}.243$  & $J = 8.048\pm 0.023$ & $H = 8.067\pm 0.031$ & $K_{s} = 8.009\pm 0.029$\\
DENIS J113106.9--625649  & $0^{\prime\prime}.279$  & $I = 8.977\pm 0.04$  & $J = 8.154\pm 0.05$  & $K_{s} = 8.111\pm 0.05$\\
USNO-B1.0 0270--0309619  & $0^{\prime\prime}.246$  & $B = 8.2\pm 0.3$     & $R = 8.2\pm 0.3$     & $I = 8.2\pm 0.3$\\ \hline
\multicolumn{5}{c}{IGR J11435--6109/CXOU J114400.3--610736}\\ \hline
2MASS J11440030--6107364 & $0^{\prime\prime}.082$ & $J = 13.003\pm 0.022$ & $H = 12.338\pm 0.021$ & $K_{s} = 11.852\pm 0.019$\\
DENIS J114400.2--610736  & $0^{\prime\prime}.397$ & $I = 14.507\pm 0.03$  & $J = 13.019\pm 0.08$  & $K_{s} = 11.808\pm 0.10$\\
USNO-B1.0 0288--0337502  & $0^{\prime\prime}.494$ & $B = 16.6\pm 0.3$     & $R = 15.7\pm 0.3$     & $I = 14.8\pm 0.3$\\ \hline
\multicolumn{5}{c}{IGR J14515--5542/CXOU J145133.1--554038}\\ \hline
2MASX J14513316--5540388 & $0^{\prime\prime}.140$ & $J = 11.142\pm 0.066$ & $H = 10.045\pm 0.045$ & $K_{s} = 9.736\pm 0.050$\\
USNO-B1.0 0343--0472749  & $0^{\prime\prime}.195$ & $B = 16.3\pm 0.3$ & $R = 12.7\pm 0.3$ & $I = 12.8\pm 0.3$\\ \hline
\multicolumn{5}{c}{IGR J17200--3116/CXOU J172005.9--311659}\\ \hline
2MASS J17200591--3116596 & $0^{\prime\prime}.269$ & $J = 13.581\pm 0.056$ & $H = 12.334\pm 0.057$ & $K_{s} = 11.983\pm 0.043$\\
DENIS J172005.9--311659  & $0^{\prime\prime}.353$ & $I = 16.212\pm 0.07$  & $J = 13.553\pm 0.09$  & $K_{s} = 11.938\pm 0.08$\\ \hline
\multicolumn{5}{c}{IGR J18214--1318/CXOU J182119.7--131838}\\ \hline
2MASS J18211975--1318388 & $0^{\prime\prime}.075$ & $J = 12.786\pm 0.052$ & $H = 11.657\pm 0.099$ & $K_{s} = 10.359$\\
DENIS J182119.7--131838  & $0^{\prime\prime}.255$ & $I = 17.007\pm 0.11$  & $J = 12.757\pm 0.12$ &  $K_{s} = 10.639\pm 0.14$\\
USNO-B1.0 0766--0475700  & $0^{\prime\prime}.574$ & ---                   & $R = 20.3\pm 0.3$    &  $I = 17.0\pm 0.3$\\ \hline
\multicolumn{5}{c}{IGR J18256--1035/CXOU J182543.8--103501}\\ \hline
2MASS J18254409--1035059 & $5^{\prime\prime}.698$ & $J = 15.015\pm 0.091$ & $H = 13.935\pm 0.077$ & $K_{s} = 13.520\pm 0.078$\\ \hline
\multicolumn{5}{c}{IGR J18259--0706/CXOU J182557.5--071022}\\ \hline
2MASS J18255759--0710229 & $0^{\prime\prime}.279$ & $J = 15.213\pm 0.151$ & $H = 13.900\pm 0.151$ & $K_{s} = 12.994\pm 0.093$\\ \hline
\multicolumn{5}{c}{IGR J18325--0756/CXOU J183228.3--075641}\\ \hline
2MASS J18322828--0756420 & $0^{\prime\prime}.625$ & $J = 16.398\pm 0.177$ & $H = 15.267\pm 0.131$ & $K_{s} = 14.490$\\ \hline
\end{tabular}
\end{minipage}
\end{table}

\begin{table}
\caption{Summary of Results for Sources with {\em Chandra} Counterparts\label{tab:summary}}
\begin{minipage}{\linewidth}
\footnotesize
\begin{tabular}{lcccccl} \hline \hline
IGR Name     & $l$\footnote{Galactic longitude in degrees.} & $b$\footnote{Galactic latitude in degrees.} & $J$-band  & Source & Spectral & $N_{\rm H,local}$\footnote{The estimate for the column density due to material local to the source based on a calculation of the measured $N_{\rm H}$ minus the Galactic $N_{\rm H}$ minus two times the Galactic $N_{\rm H_{2}}$.  An upper limit indicates no evidence for local absorption, and a range indicates evidence for slight or significant local absorption.}\\
             &                                              &                                             & Magnitude & Type\footnote{CV = Cataclysmic Variable, HMXB = High-Mass X-ray Binary, AGN = Active Galactic Nucleus.}   & Type     &  ($10^{22}$ cm$^{-2}$)\\ \hline
\multicolumn{7}{c}{{\em Chandra} cycle 8}\\ \hline
J00234+6141  & 119.62 & --1.00 & $15.12\pm 0.05$ & CV     & Low Mass  & $<$0.22\\
J06074+2205  & 188.39 &  +0.80 & $10.49\pm 0.02$ & HMXB   & Be        & 5--10\\
J09026--4812 & 268.88 & --1.09 & $15.57\pm 0.08$ & HMXB?  & ?         & $<$2.5\\
J10101--5645 & 282.24 & --0.67 & $12.62\pm 0.04$ & HMXB   & Early Giant     & $<$4.4\\
J11305--6256 & 293.87 & --1.49 & $8.05\pm 0.02$  & HMXB   & B0IIIe    & $<$0.60\\
J11435--6109 & 296.05 &  +0.97 & $13.00\pm 0.02$ & HMXB   & Be        & 10--20\\
J14515--5542 & 319.34 &  +3.29 & $11.14\pm 0.07$ & AGN    & Seyfert 2 & 0.2--1.4\\
J17200--3116 & 355.02 &  +3.35 & $13.58\pm 0.06$ & HMXB   & High Mass & 0.8--2.5\\
J18214--1318 & 17.73  &  +0.47 & $12.79\pm 0.05$ & HMXB?  & ?         & 7--15\\
J18256--1035 & 20.58  &  +0.84 & $>$15.0         & ?      & ?         & 0.1--4.7\\
J18259--0706 & 23.64  &  +2.35 & $15.21\pm 0.15$ & AGN    & Seyfert 1 & $<$1.9\\
J18325--0756 & 23.72  &  +0.56 & $16.26\pm 0.04$ & HMXB?  & ?         & 6--78\\ \hline
\multicolumn{7}{c}{{\em Chandra} cycle 6}\\ \hline
J16167--4957\footnote{We obtained {\em Chandra} observations for these 4 sources in cycle 6, and we reported the results in \cite{tomsick06}.  We include these here to allow for comparison to the cycle 8 results, which are the focus of this paper.  For 3 of these sources, the source type has been confirmed spectroscopically \citep{masetti06v}.} & 333.06 & +0.50 & $14.86\pm 0.06$ & CV & Low Mass & $<$0.8\\
J16195--4945$^{d}$ & 333.56 & +0.34  & $13.57\pm 0.03$ & HMXB? & ? & 1.1--12\\
J16207--5129$^{d}$ & 332.46 & --1.05 & $10.44\pm 0.02$ & HMXB & Early Supergiant & $<$5.1\\
J17195--4100$^{d}$ & 326.98 & --2.14 & $14.1\pm 0.1$ & CV & Low Mass & $<$0.21\\ \hline
\end{tabular}
\end{minipage}
\end{table}

\begin{table}
\caption{Summary of Results for 8 Sources without Clear {\em Chandra} Counterparts\label{tab:summary2}}
\begin{minipage}{\linewidth}
\footnotesize
\begin{tabular}{lccccc} \hline \hline
IGR Name     & $l$\footnote{Galactic longitude in degrees.} & $b$\footnote{Galactic latitude in degrees.} & {\em Chandra} Flux &  Number of & Source\\
             &                                              &                                             & Upper Limit\footnote{Upper limit on the 0.3--10~keV unabsorbed X-ray flux in units of ergs~cm$^{-2}$~s$^{-1}$.  In each case, the spectral shape assumed to calculate the upper limit is an absorbed power-law with the column density at the Galactic value.  The range of upper limits correspond to a range of assumed power-law photon indeces between $\Gamma = 1$ and 2.} & Sources Detected\footnote{The number of {\em Chandra} sources detected in the 90\% confidence {\em INTEGRAL} error circle.} & Type\footnote{HMXB = High-Mass X-ray Binary, BHC = Black Hole Candidate, SFXT = Supergiant Fast X-ray Transient.}\\ \hline
J01363+6610  & 127.45 & +3.70  & $<$(3.2--4.1)$\times 10^{-14}$ & 3 & HMXB? Be?\\
J17285--2922 & 357.64 & +2.88  & $<$(5.5--6.4)$\times 10^{-14}$ & 6 & BHC?\\
J17331--2406 & 2.61   & +4.93  & $<$(1.0--1.5)$\times 10^{-14}$ & 0 & BHC?\\
J17407--2808 & 0.10   & +1.34  & $<$(6.0--7.2)$\times 10^{-14}$ & 7 & SFXT?\\
J17445--2747 & 0.86   & +0.81  & $<$(4.1--4.5)$\times 10^{-14}$ & 2 & ?\\
J17507--2856 & 0.58   & --0.96 & $<$(7.4--8.4)$\times 10^{-14}$ & 3 & ?\\
J18193--2542 & 6.51   & --4.91 & $<$(0.9--1.4)$\times 10^{-14}$ & 0 & ?\\
J18539+0727  & 39.85  & +2.85  & $<$(1.7--2.0)$\times 10^{-14}$ & 0 & BHC\\ \hline
\end{tabular}
\end{minipage}
\end{table}


\clearpage

\end{document}